\documentclass[aps,prb,reprint,amsfonts,amsmath,amssymb,showpacs,groupedaddress,superscriptaddress]{revtex4-1}
\usepackage{graphicx}
\usepackage{bm}
\usepackage{color}
\usepackage{subfigure}
\usepackage[colorlinks,urlcolor=blue,linkcolor=blue,anchorcolor=blue,citecolor=blue,bookmarks]{hyperref}
\usepackage{ulem}

\begin{document}

\title{%Dynamical properties and fractionalized excitations of the antiferromagnetic $J_{1}$-$J_{2}$ Heisenberg model on the honeycomb lattice
 Spin dynamics and continuum spectra of the honeycomb $J_{1}$-$J_{2}$  antiferromagnetic Heisenberg model}

\author{Cheng Gu}
\affiliation{National Laboratory of Solid State Microstructures and School of Physics, Nanjing University, Nanjing 210093, China}

\author{Shun-Li Yu}
\email{slyu@nju.edu.cn}
\affiliation{National Laboratory of Solid State Microstructures and School of Physics, Nanjing University, Nanjing 210093, China}
\affiliation{Collaborative Innovation Center of Advanced Microstructures, Nanjing University, Nanjing 210093, China}

\author{Jian-Xin Li}
\email{jxli@nju.edu.cn}
\affiliation{National Laboratory of Solid State Microstructures and School of Physics, Nanjing University, Nanjing 210093, China}
\affiliation{Collaborative Innovation Center of Advanced Microstructures, Nanjing University, Nanjing 210093, China}

\date{\today}

\begin{abstract}
  We employ the spin cluster perturbation theory to investigate the dynamical properties of the antiferromagnetic $J_{1}$-$J_{2}$ Heisenberg model on the honeycomb lattice. We obtain the excitation spectra for all possible phases in the phase diagram, including the N\'{e}el phase, plaquette valence-bond-solid phase, dimer valence-bond-solid phase and stripe antiferromagnetic phase. In the N\'{e}el phase, besides the obvious renormalization of the magnon dispersion, we find that the spectrum exhibits a dome-shaped broad continuum around the second Brillouin zone (BZ) and the additional strong continuum close to the corner of the BZ. In the valence-bond-solid phases, the spectra are dominated by a strong broad continuum all the way down to below $J_1$ coexisting with the lowest-energy triplon modes characterizing the plaquette and dimer phases. We ascribe this strong broad continuum and the additional continuum close to the BZ corner in the N\'{e}el phase to the contributions of fractionalized spinon excitations. In the stripe phase, a clear difference from the linear spin wave approximation is that the spectrum is gapped at the $M$ point while that obtained by the latter is gapless due to the strong quantum fluctuations. We point out that the features observed in the N\'{e}el phase are consistent with the recent neutron scattering experiments on YbCl$_{3}$ and YbBr$_{3}$.
\end{abstract}

\maketitle

\section{\label{sec:SectionI}Introduction}

The low-dimensional $s=1/2$ antiferromagnetic (AF) Heisenberg model is the fundamental model in the studies of quantum magnetism. By changing the geometry of the underlying lattice and the range of the exchange coupling to introduce frustration, the spin interactions between magnetic degrees of freedom could be incompatible with the underlying lattice geometry and thus exotic ground states and excitations can emerge\cite{PhysRevB.39.2344,RevModPhys.63.1,sachdev2008,Savary_2016,Broholmeaay0668,RevModPhys.89.025003}. A typical example is the square-lattice $J_{1}$-$J_{2}$ Heisenberg model, in which the $J_{1}$ and $J_{2}$ terms denote the nearest-neighbor (NN) and next-nearest-neighbor (NNN) exchange interactions, respectively. By tuning the magnitudes of the interactions, this model can realize not only the magnetically ordered phases with N\'{e}el order and stripe order in the weak and strong frustration region respectively, but also the magnetically disordered phases in the intermediate frustration region ($0.4\lesssim J_{2}/J_{1}\lesssim0.6$) \cite{PhysRevLett.87.097201,PhysRevB.73.184420,PhysRevB.86.024424,PhysRevLett.111.037202,
PhysRevLett.113.027201,PhysRevB.96.014414,PhysRevB.97.174408,PhysRevLett.121.107202,PhysRevX.11.031034}, which have been ascribed to be the quantum spin liquid (QSL) or valence bond solid (VBS) phases. Besides the ground state phase diagram, the spin dynamics is also crucial for understanding the rich physics of this quantum spin system and has attracted considerable interest recently in both experimental and theoretical aspects. Especially, a recent inelastic neutron scattering (INS) measurement on the Cu(DCOO)$_{2}\cdot$4D$_{2}$O which is considered as the best realization of the square-lattice Heisenberg antiferromagnet reveals that in addition to the well-defined low-energy magnon excitations, there is an obvious high-energy continuum at $(\pi,0)$ in the Brillouin zone (BZ)\cite{Naturephys2015}. The similar anomalous high-energy continuum at $(\pi, 0)$ is also observed in the parent antiferromagnet La$_2$CuO$_4$ of high-T$_c$ cuprates\cite{PhysRevLett.105.247001}. Although the nature of this anomalous continuum is still under debate \cite{Naturephys2015,PhysRevLett.105.247001,PhysRevX.7.041072,PhysRevB.52.R15695,PhysRevLett.86.528,SciPostPhys.4.1.001,PhysRevB.98.100405,PhysRevB.98.134410,PhysRevB.99.094412,PhysRevB.71.184440,PhysRevLett.86.5377,2007Quantumdynamics}, it suggests the possibility for a coexistence of conventional magnon excitations and deconfined spinons which are the spin-$1/2$ excitations by fractionalizing the spin-$1$ magnons.

The existence of non-spin-wave excitations in the square-lattice N\'{e}el antiferromagnet indicates that the magnetically ordered state of the AF Heisenberg model has noticeable quantum fluctuations. From a theoretical point of view, the quantum fluctuations in the honeycomb-lattice AF Heisenberg model can be further enhanced due to its low coordination number, so we can expect more obvious characteristics of collective quantum behaviors beyond the magnon in the magnetically ordered phases. Furthermore, the very recent INS experiment on YbCl$_{3}$ \cite{2021Van}, which is suggested to be a realization of the ideal NN antiferromagnetic Heisenberg model on the honeycomb lattice, has shown a conventional magnon mode and a dome-shaped broad continuum in the spin excitation spectra which has been suggested to come from two-magnon excitations due to longitudinal spin fluctuations. Interestingly, we also notice that a ball of particularly high spectral weights exhibiting as a additional continuum superimposed on the dome-shaped continuum  exists around the corners ($K$ point) of the BZ and disappears near the center of the BZ, and this part of the continuum spectra within a limited region exhibits a deviation from the two-magnon continuum. Similarly, for YbBr$_{3}$ with the same structure as YbCl$_{3}$, the recent INS measurement also reveals a broad high-energy continuum at the boundary of BZ in addition to well-defined $S=1$ magnetic excitations near the zone center \cite{npjqm.5.85}, while the experiment only observes a short-range magnetic order. Thus, the AF Heisenberg model on the honeycomb lattice provides another ideal platform for studying the anomalous excitations due to quantum fluctuations in the AF ordered magnets.

Like its counterpart on the square lattice, the phase diagram of the honeycomb-lattice $J_{1}$-$J_{2}$ Heisenberg model has also been extensively studied by different theoretical methods, such as the linear-spin-wave (LSW) and bond-operator theory \cite{PhysRevB.81.214419}, series expansions\cite{PhysRevB.84.094424}, coupled-cluster theory \cite{PhysRevB.84.012403,PhysRevB.89.220408}, modified spin wave theory \cite{2016Quantum}, variational Monte Carlo (VMC)\cite{PhysRevLett.107.087204,PhysRevB.89.094413,PhysRevB.96.104401}, exact diagonalization (ED) \cite{2001An,PhysRevB.84.024406}, and density matrix renormalization group\cite{PhysRevLett.110.127203,PhysRevLett.110.127205,PhysRevB.88.165138}. Most of these studies report a nonmagnetic phase for $J_{2}/J_{1}\sim0.2-0.4$, in which two VBS states with plaquette and staggered-dimer orders are realized, but the results are in general disagreement on the range of this phase. For $J_{2}/J_{1}<0.2$, the ground state is a AF N\'{e}el state, while for $J_{2}/J_{1}>0.4$, the results based on the ED \cite{PhysRevB.84.024406}, VMC\cite{PhysRevB.89.094413}, modified spin wave theory \cite{2016Quantum} and coupled-cluster theory \cite{PhysRevB.89.220408} predicted a magnetic phase with stripe order. By contrast, the studies of the spin dynamics of this model obviously lag behind that of the ground-state phase diagram. So far, the spin excitation spectrum beyond the LSW theory for this model comes from the VMC study for the N\'{e}el phase and two possible valence bond solid phases with $J_{2}\leq0.4J_{1}$ \cite{ferrari2020dynamical}, as well as the random phase approximation based on the resonant-valence-bond (RVB) ansatz with the AF order\cite{PhysRevB.104.L180406} and the Schwinger boson mean-field theory\cite{PhysRevB.97.205112}. Therefore, to fully explore the spin dynamics of this model and provide theoretical understandings on the  related recent experimental observations, we need a more comprehensive research.

In this paper, we study the spin dynamics of the AF $J_{1}$-$J_{2}$ Heisenberg model on the honeycomb lattice by using the spin cluster perturbation theory (CPT) \cite{PhysRevB.98.134410}. We obtain the excitation spectra for all possible phases in the phase diagram, including the N\'{e}el phase, plaquette valence-bond-solid phase, dimer valence-bond-solid phase and stripe antiferromagnetic phase. In the N\'{e}el phase with $J_{2}/J_{1}=0.1$, we reproduce well the entire spectrum observed in recent INS experiment on YbCl$_{3}$, especially the strong continuum close to the corner of the BZ, where our results reveal that there is even no well defined one-magnon mode. The continuum around the BZ corner had not been analysed in the previous VMC study~\cite{ferrari2020dynamical} and has also been shown here to reproduce the experimental observations in YbBr$_{3}$~\cite{npjqm.5.85}. Furthermore, We find that this continuum is inconsistent with the multi-magnon mechanics, and suggest that it originates from the deconfinement of fractionalized spin-$1/2$ spinons based on a further analysis of the evolution of the continuum from the N\'{e}el phase to the plaquette VBS phase. In the valence-bond-solid phases, the spectra show a clear gap and a broad continuum, and different triplon excitation dispersions at lowest energies corresponding to plaquette and dimer valence-bond-solid phases, respectively. Finally, we go beyond the previous studies~\cite{ferrari2020dynamical,PhysRevB.104.L180406,PhysRevB.97.205112} to explore the properties of spin excitations in the large $J_{2}$ regime where a stripe AF phase is expected, we find that the quantum fluctuations not only play an important role in stabilizing the stripe order of the ground state, but also lead to the obvious deviation of the excitation spectrum from the results of the LSW theory.

The paper is organized as follows. In Sec.\ref{sec:SectionII}, we introduce the model and the spin cluster perturbation theory (CPT). In Sec.\ref{sec:SectionIII}, we present the dynamic excitation spectra for various possible phases of the $J_{1}$-$J_{2}$ Heisenberg model on the honeycomb lattice. Section \ref{sec:SectionVI} presents a summary.

\section{\label{sec:SectionII}Model and Method}

\begin{figure}
  \includegraphics[width=\columnwidth]{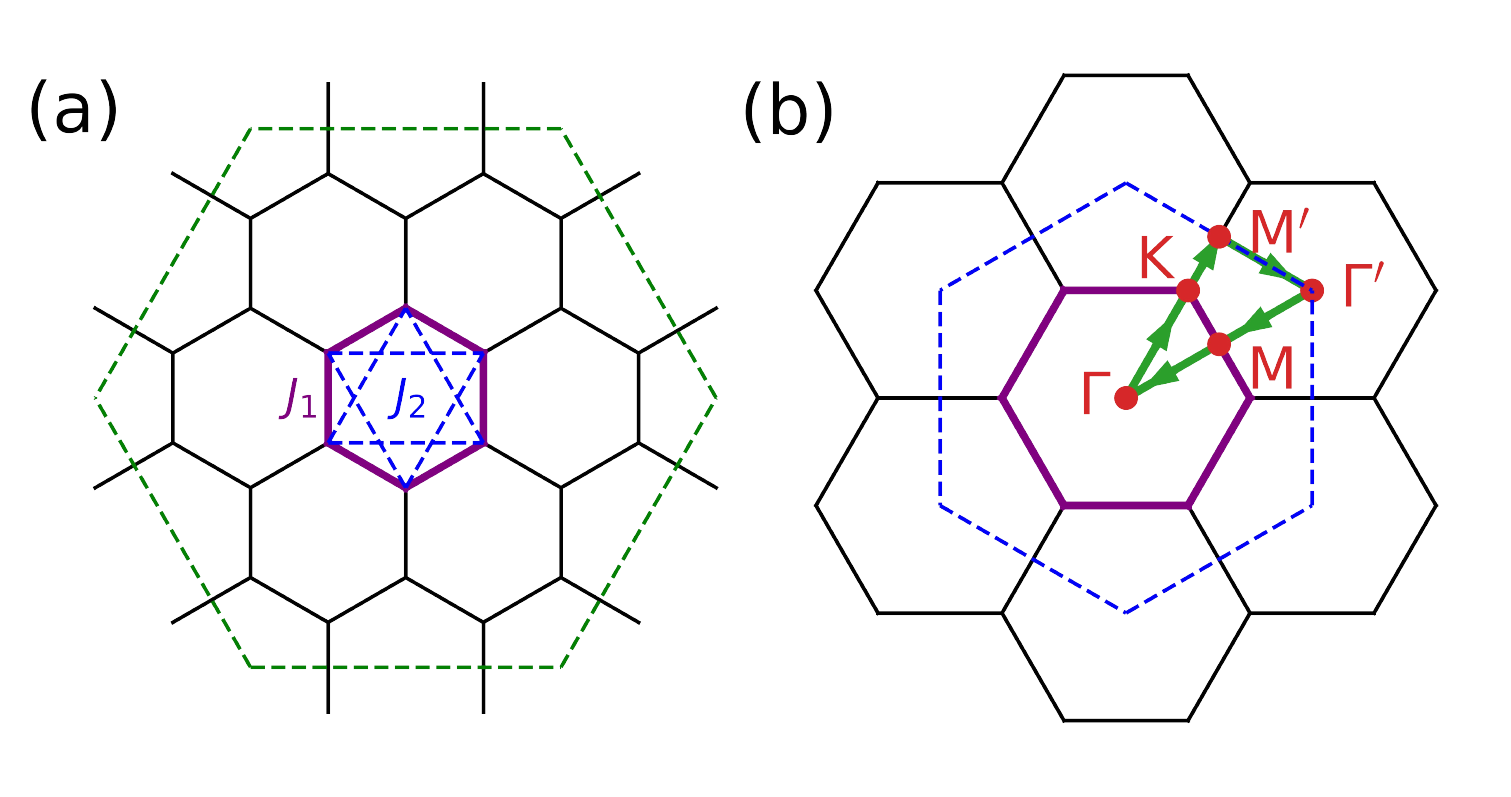}
  \caption{\label{fig:lattice}(Color online) (a) 24-site cluster used in the CPT calculations to tile the honeycomb lattice. (b) Path consisting of the high symmetry lines in the BZ used to illustrate the excitation spectra in this paper unless otherwise specified.}
\end{figure}

The $J_{1}$-$J_{2}$ Heisenberg model on the honeycomb lattice is given by
\begin{equation}
    H=J_1\sum_{\langle ij \rangle} \bm{S}_i \bm{\cdot} \bm{S}_j + J_2\sum_{\ll ij \gg} \bm{S}_i \bm{\cdot} \bm{S}_j, \label{eq:Hamiltonian}
\end{equation}
where $\langle ij \rangle$ denotes the NN bonds, $\ll ij \gg$ the NNN bonds. As mentioned above, the ground-state phase diagram of this model has been extensively explored previously\cite{PhysRevB.81.214419,PhysRevB.84.094424,PhysRevB.84.012403,PhysRevB.89.220408,2016Quantum,PhysRevLett.107.087204,
PhysRevB.89.094413,PhysRevB.96.104401,2001An,PhysRevB.84.024406,PhysRevLett.110.127203,PhysRevLett.110.127205,PhysRevB.88.165138}, so here we will focus on the spin dynamical properties.

In order to obtain reliable spin excitation spectrum with high momentum resolution, we use the spin cluster perturbation theory (CPT), which has been successfully applied to the $J_{1}$-$J_{2}$ Heisenberg model on the square lattice \cite{PhysRevB.98.134410}. Normally, the CPT is used to study the charge dynamics of the Hubbard model\cite{Phys.Rev.Lett.84.522,PhysRevLett.85.2585,PhysRevLett.92.126401,PhysRevLett.107.010401,PhysRevB.84.064520,PhysRevB.85.144402}. To extend the CPT method to spin systems, we use the mapping between spin-$1/2$ operators and hard-core bosonic operators \cite{ProgTheorPhys.16.569,Sov.Phys.JETP.60.781}:
\begin{equation}
    S_i^{+} = b_i^{\dagger}, \quad
    S_i^{-} = b_i, \quad
    S_i^{z} = b_i^{\dagger} b_i - \frac{1}{2},  \label{eq:Hard-core boson}
\end{equation}
where $b_i^{\dagger}$ and $b_i$ are the creation and annihilation operators of the hard-core boson, which can be viewed as particles with infinite on-site repulsion interaction. The hard-core boson operators have the following relations: $[b_{i},b_{j}]=[b^{\dag}_{i},b^{\dag}_{j}]=0$ for $i\neq j$, $(b_{i})^{2}=(b^{\dag}_{i})^{2}=0$, and $[b_{i},b^{\dag}_{j}]=\delta_{ij}(1-2b^{\dag}_{i}b_{i})$. Under these relations, the occupation number per site is restricted to $n_{i}=0$ or $1$ with $n_{i}=b^{\dag}_{i}b_{i}$, and the commutation relations of the spin operators, $[S_{i}^{+},S_{j}^{-}]=2\delta_{ij}S_i^{z}$ and $[S_{i}^{\pm},S_{j}^{z}]=\mp\delta_{ij}S_i^{\pm}$, are realized.
Using this representation, the Hamiltonian (\ref{eq:Hamiltonian}) is rewritten as
\begin{align}
H&=\frac{1}{2}J_{1}\sum_{\langle ij\rangle}(b_{i}^{\dag}b_{j}+h.c.)+\frac{1}{2}J_{2}\sum_{\langle\langle ij\rangle\rangle}(b_{i}^{\dag}b_{j}+h.c.) \nonumber \\
&+J_{1}\sum_{\langle ij\rangle}n_{i}n_{j}+J_{2}\sum_{\langle\langle ij\rangle\rangle}n_{i}n_{j}-2(J_{1}+J_{2})\sum_{i}n_{i}.
\label{eq:hard-core-boson}
\end{align}
In the CPT method, we first divide the original lattice into identical clusters to form a superlattice [see Fig.~\ref{fig:lattice}(a)], and the Hamiltonian can be consequently rewritten as $H=H_{c}+V$, where $H_{c}$ is the cluster Hamiltonian and $V$ represents the coupling between different clusters. Then, the cluster Green function $\bm{G}(z)$ (in matrix form) of the frequency $z$ is calculated by ED method at zero temperature~\cite{RevModPhys.66.763} and finite temperatures~\cite{PhysRevB.49.5065} with open boundary condition, and the original lattice Green function is given by:
\begin{equation}
    \bm{g} (\tilde{\bm{k}},z) = \bm{G}(z) [1-\bm{V}(\tilde{\bm{k}})\bm{G}(z)]^{-1}, \label{eq:CPT GF I}
\end{equation}
where $\tilde{\bm{k}}$ is the wave vector in the BZ of the superlattice and $V_{\mu\nu}(\tilde{\bm{k}})=\sum_{\bm{R}}V^{0\bm{R}}_{\mu\nu}e^{\tilde{\bm{k}}\cdot\bm{R}}$ with $\bm{R}$ the superlattice index, $\mu$ and $\nu$ the site indices in a cluster. $V^{0\bm{R}}_{\mu\nu}=\frac{1}{2}J_{1}\sum_{\bm{e}}\delta_{\bm{R}+\bm{r}_{\nu}-\bm{r}_{\mu},\bm{e}}
+\frac{1}{2}J_{2}\sum_{\bm{e}^{\prime}}\delta_{\bm{R}+\bm{r}_{\nu}-\bm{r}_{\mu},\bm{e}^{\prime}}$ contains
all hopping terms of the hard-core bosons between two clusters at $0$ and $\bm{R}$, and $\bm{e}$ and $\bm{e}^{\prime}$ denote the NN and NNN vectors, respectively. Since the inter-cluster couplings contained in $V$ should be quadratic in the CPT method, we perform the following mean-field approximation on the NN and NNN interactions in the Hamiltonian (\ref{eq:hard-core-boson}) between different clusters \cite{PhysRevB.98.134410}:
\begin{align}
J_{1}\sum_{\langle ij\rangle}(n_{i}\langle n_{j}\rangle+\langle n_{i}\rangle n_{j})
+J_{2}\sum_{\langle\langle ij\rangle\rangle}(n_{i}\langle n_{j}\rangle+\langle n_{i}\rangle n_{j}).
\end{align}
Due to the cluster decomposition, the Green function obtained by the CPT method breaks the original lattice translation symmetry. We then perform a periodization procedure to recover the translation invariance and the CPT Green function is given by
\begin{equation}
    g_{cpt} (\bm{k},z) = \frac {1}{L}\sum_{\mu,\nu} e^{-i\bm{k} \cdot {(\bm{r}_{\mu}-\bm{r}_{\nu})}}g_{\mu\nu} (\tilde{\bm{k}},z),  \label{eq:CPT GF II}
\end{equation}
where $L$ is the number of sites in each cluster. Each wave vector $\bm{k}$ in the BZ of original lattice can be expressed as $\bm{k}=\tilde{\bm{k}}+\bm{K}$, where $\bm{K}$ is the reciprocal vector of the superlattice.
The mapping between $s=1/2$ spins and hard-core bosons Eq.(2) leads to a straightforward relation of the dynamical spin susceptibility $\mathcal{S}^{+-}(\bm{k},\omega)$ to the bosonic single-particle Green function, as given by:
\begin{equation}
    \mathcal{S}^{+-}(\bm{k},\omega)=-\mathrm{Im}g_{cpt}(\bm{k},\omega +i\eta).  \label{eq:Structure Factor}
\end{equation}

In this paper, the dynamical spin susceptibilities are calculated at zero temperature by using a $24$-site cluster with $C_{6}$ symmetry as shown in Fig.~\ref{fig:lattice}(a) to divide the original lattice and setting the broadening factor $\eta=0.15J_{1}$. The 24-site cluster is the maximum cluster size we can handle preserving the $C_{6}$ rotational symmetry of the original lattice. We have checked the results by using clusters with different sizes and structures, such as for the 18-site cluster, the results are qualitatively consistent with those presented here for the 24-site cluster.

\section{\label{sec:SectionIII}Results}

\subsection{\label{sec:A} CPT results in N\'{e}el phase}
\begin{figure}
  \includegraphics[width=\columnwidth]{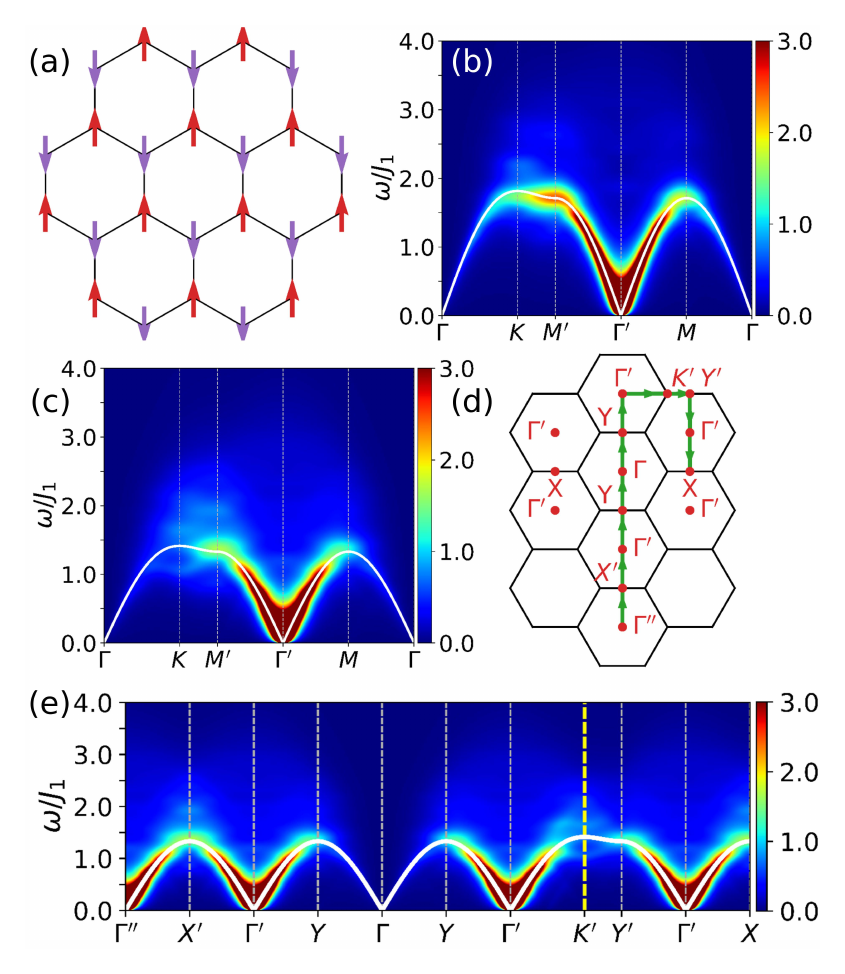}
  \caption{\label{fig:neel-order}(Color online) (a) Illustration of the pattern of the AF N\'{e}el order. (b) Dynamical structure factor $S^{+-}(\bm{k},\omega)$ for $J_{2}/J_{1}=0$. (c) $S^{+-}(\bm{k},\omega)$ for $J_{2}/J_{1}=0.1$. (d) Path for the illustration of the spectrum in (e). (e) $S^{+-}(\bm{k},\omega)$ for $J_{2}/J_{1}=0.1$ along the path shown in (d). The white lines in (b), (c) and (e) are the LSW dispersion with the $1/S$ correction.}
\end{figure}
Let us start our discussion with the unfrustrated Heisenberg model for $J_2=0$, whose ground state has an AF N\'{e}el order [see Fig.~\ref{fig:neel-order}(a)]. The spin excitation spectrum along a path consisting of the high symmetry lines [see Fig.~\ref{fig:lattice}(b)] is shown in Fig.~\ref{fig:neel-order}(b), in which we can find a well-defined magnon dispersion. The overall shape of the dispersion is consistent with that obtained by the LSW theory with $1/S$ correction~\cite{PhysRevB.82.144407}, which is shown by the white solid line. Especially, the CPT result can successfully produce the Goldstone mode at the $\Gamma^{\prime}$ point, which is a defining characteristic of the AF N\'{e}el order. According to the fact that the CPT result is consistent with the spin-wave dispersion obtained by the LSW theory, we can expect that there is also a Goldstone mode at the $\Gamma$ point, but the vanishing spectral intensity makes it hardly to be seen. However, the dispersion from $K$ to $M$ ($M^{\prime}$) obtained by the CPT method is nearly flat, which deviates significantly from the LSW dispersion with the $1/S$ correction, in which the energy at the $K$ point is obviously larger than that at the $M$ and $M^{\prime}$ points. We also note that this flat dispersion is in good agreement with the result of the VMC calculation~\cite{ferrari2020dynamical} (see appendix~\ref{appendix-a}).
The difference between the CPT and LSW dispersions is much larger than that in the square-lattice Heisenberg model \cite{PhysRevB.98.134410}, evidencing a much stronger renormalization effect due to the enhanced quantum fluctuations resulting from the low coordination number in the honeycomb lattice. Furthermore, there is a continuum in a wide energy range above the top of the magnon band, and it is especially obvious close to the $K$ point.

We further study the excitation spectrum by increasing $J_2$, and the result for the $J_2 = 0.1J_1$ is presented
in Fig.~\ref{fig:neel-order}(c). It is found that the low-energy one-magnon excitations can still be described by the LSW theory with $1/S$ correction. While, the broad continuum becomes more obvious, so that a dome-shaped region with the upper boundary centered on the $\Gamma^{\prime}$ point can now be seen clearly. More importantly, the spectral weights around the $K$ point have been suppressed heavily and no well-defined magnon mode can even be identified anymore. Thus, the whole spectrum around this point becomes completely continuum and extends out of the dome-shaped continuum reaching near $2.5J$. Thus, this continuum can be distinguished as that superimposing on the dome-shaped part which extends to be about $4.0J$ around the $\Gamma^{\prime}$ point. At the bottom of this continuum, a dim roton-like excitation with a minimum at $K$ point can also be noticed.

\subsection{\label{sec:B} CPT results in nonmagnetic phase}

\begin{figure}
  \includegraphics[width=\columnwidth]{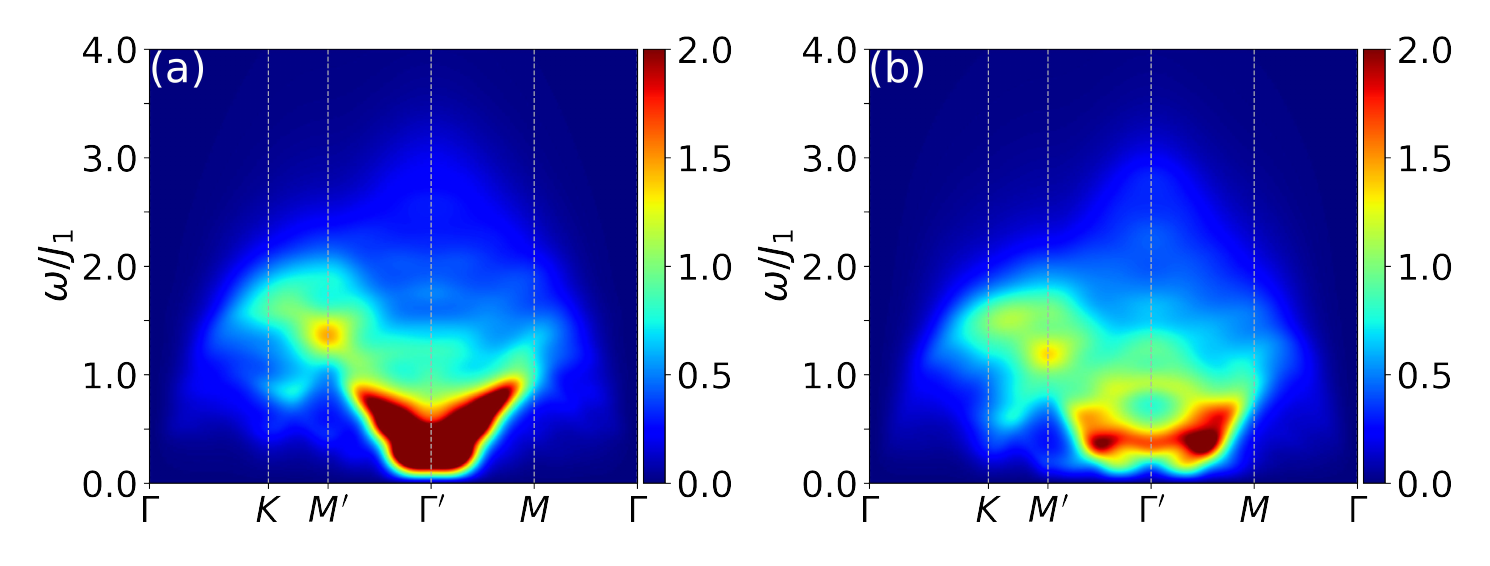}
  \caption{\label{fig:vbs}(Color online) Dynamical structure factors $S^{+-}(\bm{k},\omega)$ in the nonmagnetic phase. (a) $J_{2}/J_{1}=0.25$. (b) $J_{2}/J_{1}=0.37$.}
\end{figure}
When $J_{2}$ is further increased to be $0.18J_{1}$, we find that the AF N\'{e}el state is no longer stable and the system becomes a nonmagnetic state. This evolution of the ground states with $J_{2}$ allows us to investigate the corresponding evolution of excitation spectra, especially the strong continuum around the $K$ point in the Ne\'{e}l phase to trace its possible origin. Our numerical calculation reveals that the nonmagnetic phase persists for $0.18J_{1}<J_{2}<0.48J_{1}$, which is consistent with the previous researches in the literatures \cite{PhysRevB.81.214419,PhysRevB.84.094424,PhysRevB.84.012403,PhysRevB.89.220408,2016Quantum,PhysRevLett.107.087204,
PhysRevB.89.094413,PhysRevB.96.104401,2001An,PhysRevB.84.024406,PhysRevLett.110.127203,PhysRevLett.110.127205,PhysRevB.88.165138}.
In Fig.~\ref{fig:vbs}(a) and (b), we present the excitation spectrum based on our CPT calculation for $J_{2}=0.25J_{1}$ and $0.37J_{1}$, respectively. The common features are that the Goldstone mode disappears, the spectrum develops a clear gap and a broad continuum reaching down to below $J_{1}$.
The former two features strongly suggest that the long-range N\'{e}el order disappears and the system enters into a nonmagnetic phase. On the other hand,
they exhibit different structures at the lowest excitation energies. For $J_{2}=0.25J_{1}$, the spectral bottom is nearly flat with the minimum at the $\Gamma^{\prime}$ point  [Fig.~\ref{fig:vbs}(a)]. While for $J_{2}=0.37J_{1}$, the lowest-energy spectrum is split into two minima and they moves towards the $M$ point [Fig.~\ref{fig:vbs}(b)]. These salient different features clearly suggest that two different phases exist in these two parameter regions. We notice that, according to the previous studies\cite{PhysRevB.81.214419,PhysRevB.84.094424,PhysRevB.84.012403,PhysRevB.89.220408,2016Quantum,PhysRevLett.107.087204,
PhysRevB.89.094413,PhysRevB.96.104401,2001An,PhysRevB.84.024406,PhysRevLett.110.127203,PhysRevLett.110.127205,PhysRevB.88.165138,PhysRevB.98.174421},
there are two VBS phases in this nonmagnetic region with a plaquette order for $J_{2}<0.36J_{1}$ and a column dimer order for $J_{2}>0.36J_{1}$, respectively. Thus, the two different excitation spectra can be naturally interpreted as those out of the plaquette VBS and dimer VBS states, respectively. The lowest-energy spectra manifest the dispersion of the well-defined triplon mode, and are consistent with VMC calculation based on the plaquette and dimer VBS variational states
\cite{ferrari2020dynamical} (see appendix~\ref{appendix-a}), though the bottom of spectrum for $J_{2}=0.37J_{1}$ is not right at the $M$ point due to the finite-size effects of the cluster we can chosen here.

\subsection{\label{sec:C} Possible origin of the continuum in N\'{e}el phase}
\begin{figure}
  \includegraphics[width=\columnwidth]{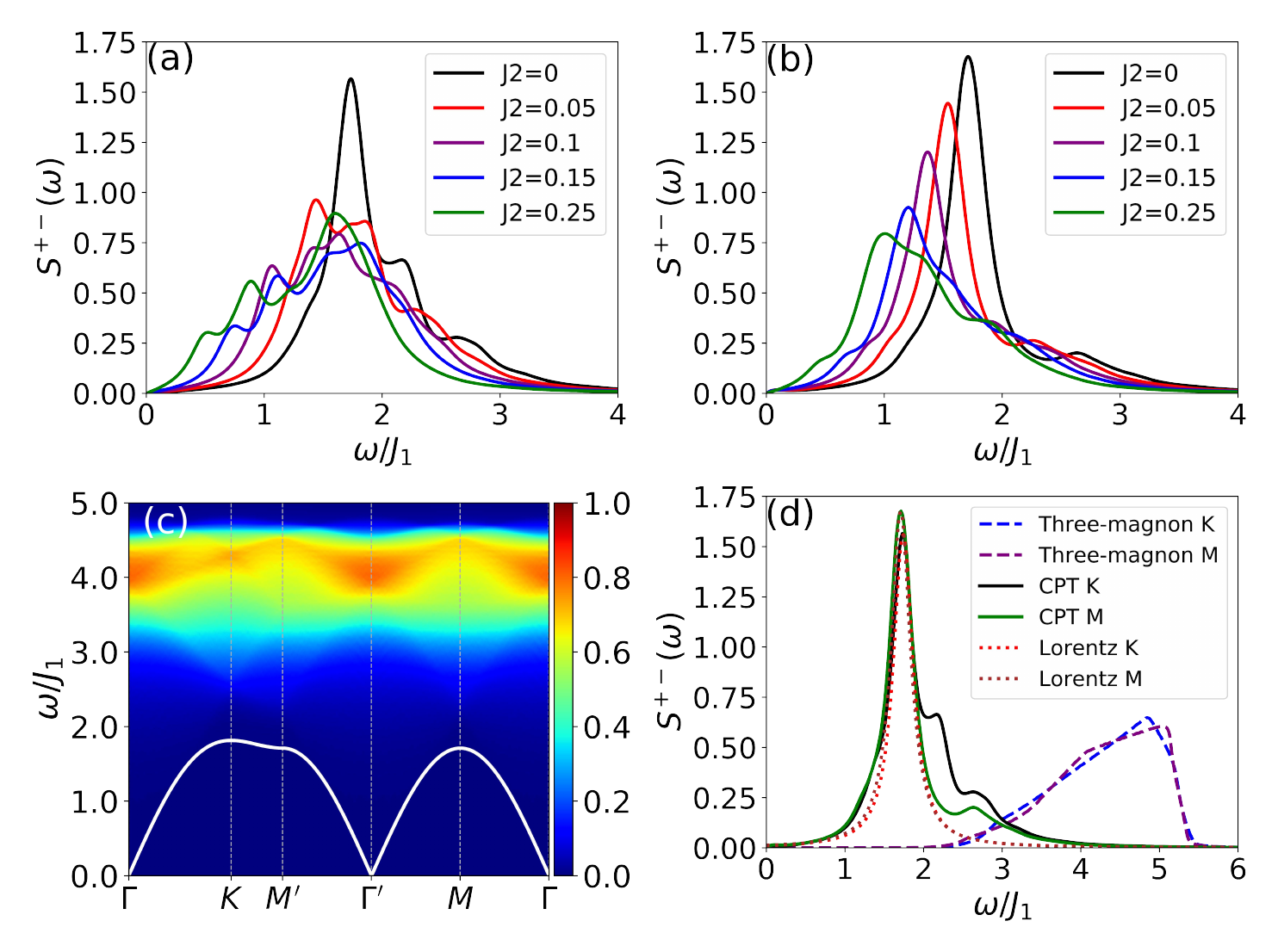}
  \caption{\label{fig:evolution}(Color online) (a) and (b) Evolution of the dynamical structure factors with $J_{2}/J_{1}$ at the $K$ and $M$ points of BZ, respectively. (c) Density of three-magnon states for the AF N\'{e}el order with $J_{2}/J_{1}=0$. The white line is the LSW dispersion with the $1/S$ correction. (d) Dynamical structure factors $S^{+-}(\bm{k},\omega)$, Lorentz fittings and densities of three-magnon states at the $K$ and $M$ points of BZ for $J_{2}/J_{1}=0$. }
\end{figure}

Here, we will focus on the discussion of the possible origin of the additional strong continuum around the $K$ point and along the $K-M$ line in the N\'{e}el phase. In order to get more insight, we study the evolution of the spectra with $J_{2}$ from the N\'{e}el phase to the plaquette VBS phase at two typical momenta $K$ and $M$ as shown in Fig.~\ref{fig:evolution}(a) and (b), respectively.
It can be seen that there is a sharp peak at both momenta for $J_{2}=0$, signifying the existence of the well-defined magnons. But, the spectrum is not exhausted completely by a Lorentz fitting [see the fitting in Fig.~\ref{fig:evolution}(d)], the lineshape develops a noticeable tail in the high-energy region. This non-Lorentz tail constitutes the continuum above the magnon dispersion around the $K$ and $M$ points, as shown in the intensity map of the spectra in Fig.~\ref{fig:neel-order}(b). From the results for $J_{2}=0$, we can also see that the spectral weight of the tail at the $K$ point is clearly larger than that at the $M$ point, so it is more easier to notice the continuum around the $K$ point as discussed above. With the increase of $J_{2}$, the sharp peak at the $K$ point disappears rapidly, such as for $J_{2}\geq 0.05$, so no well-defined magnon mode exists. Instead, a local broad peak emerges near the edge of the lineshape and corresponds to the roton-like mode identified already based on Fig.~\ref{fig:neel-order}(c). In contrast, the peak corresponding to the magnon mode at the $M$ point remains in the whole N\'{e}el phase, and a peak broadened by the continuum can also be identified in the plaquette VBS state for $J_{2}=0.25$ which is believed to come from the triplon mode. Importantly, the evolutions of the high-energy tails in the spectra at both $K$ and $M$ points are continuous from $J_{2}=0$ in the N\'{e}el phase to $J_{2}=0.25J_{1}$ in the plaquette VBS phase, in particular the spectrum at the $K$ point for $J_{2}=0.1J_{1}$ is already qualitatively consistent with that for $J_{2}=0.25J_{1}$.  It has been demonstrated that the transition from the  N\'{e}el state to the plaquette VBS state is a continuous phase transition with a deconfined quantum critical point \cite{PhysRevLett.110.127203,PhysRevLett.110.127205}, at which the elementary spin excitations are deconfined spinons \cite{PhysRevB.70.144407,Senthil2004}. Therefore, the broad continua above the lowest-energy triplon excitations we observe in the VBS states may come from the nearly-deconfined spinon excitations.
The continuous evolution of the spectra with $J_2$ imply that the additional strong continuum superimposed on the dome-shaped continuum near the $K$ point and along the $K-M$ line in the AF N\'{e}el phase may also originate from the effect of the nearly-deconfined spinons. The similar scenario that one magnon is fractionalized into two nearly free spinons for the high-energy excitations in the AF N\'{e}el phase has also been applied to interpret the high-energy continuum at $(\pi,0)$ in the square-lattice AF Heisenberg model in previous studies \cite{Naturephys2015,PhysRevX.7.041072,PhysRevB.98.100405,PhysRevB.98.134410}.

In principle, the multimagnon processes \cite{RevModPhys.85.219}, i.e., one magnon decays into multi magnons due to magnon-magnon interactions, can also lead to the continuum. We wish to comment on this possibility. In the multi-magnon scenario, for a collinear antiferromagnet as discussed here, the one-magnon states are odd parity under the $\pi$ rotation about the N\'{e}el vector direction, but the two-magnon states are even parity under this symmetry operation, so the coupling between one- and two-magnon sectors is forbidden in the Heisenberg model \cite{RevModPhys.85.219}. For this reason, the lowest-order decay channel beyond the LSW approximation is that one magnon decays into three magnons due to the four-magnon interaction \cite{RevModPhys.85.219}
\begin{align}
H_{4}=\sum_{k_{1}-k_{4}}V_{1234}a^{\dag}_{k_{1}}a^{\dag}_{k_{2}}a^{\dag}_{k_{3}}a_{k_{4}}\delta(k_{1}+k_{2}+k_{3}-k_{4}).
\end{align}
The effect of such terms on the continuum depends crucially on the availability of low-energy three-magnon states. To gain insight into whether the multi-magnon mechanism can fully reproduce the features of the continuum, we check the density of three-magnon states, which is given by \cite{RevModPhys.85.219,Nat.Commun.8.1152}
\begin{align}
D(k)=\frac{1}{N^{2}}\sum_{p,q}\delta(\epsilon_{k}-\epsilon_{p}-\epsilon_{q}-\epsilon_{k-p-q}),
\end{align}
where $\epsilon_{k}$ is the magnon dispersion. The density of three-magnon states for $J_{2}/J_{1}=0$, based on the one-magnon dispersion obtained by the LSW theory with the $1/S$ correction, is shown in Fig.~\ref{fig:evolution}(c). We find that the large densities of three-magnon states are concentrated at high energies above $3J_{1}$, but the continua in Fig.~\ref{fig:neel-order}(b) are concentrated at energies below $3J_{1}$.
In order to more clearly show the energy distributions of the continuum and the three-magnon states, we present the CPT spectra together with a Lorentz fitting and the densities of three-magnon states at the $K$ and $M$ points in Fig.~\ref{fig:evolution} (d). We can see clearly that the energy distribution region of the continuum is obviously different from that of the three-magnon excitations, and the spectral weight in the small overlapping region is very weak. Therefore, we expect that the continua in the N\'{e}el state are at least not mainly due to the multimagnon processes, though we do not consider the renormalization of the magnon energy band due to the magnon-magnon interactions for simplicity.

\subsection{\label{sec:D}CPT results in stripe phase}
\begin{figure}
  \includegraphics[width=\columnwidth]{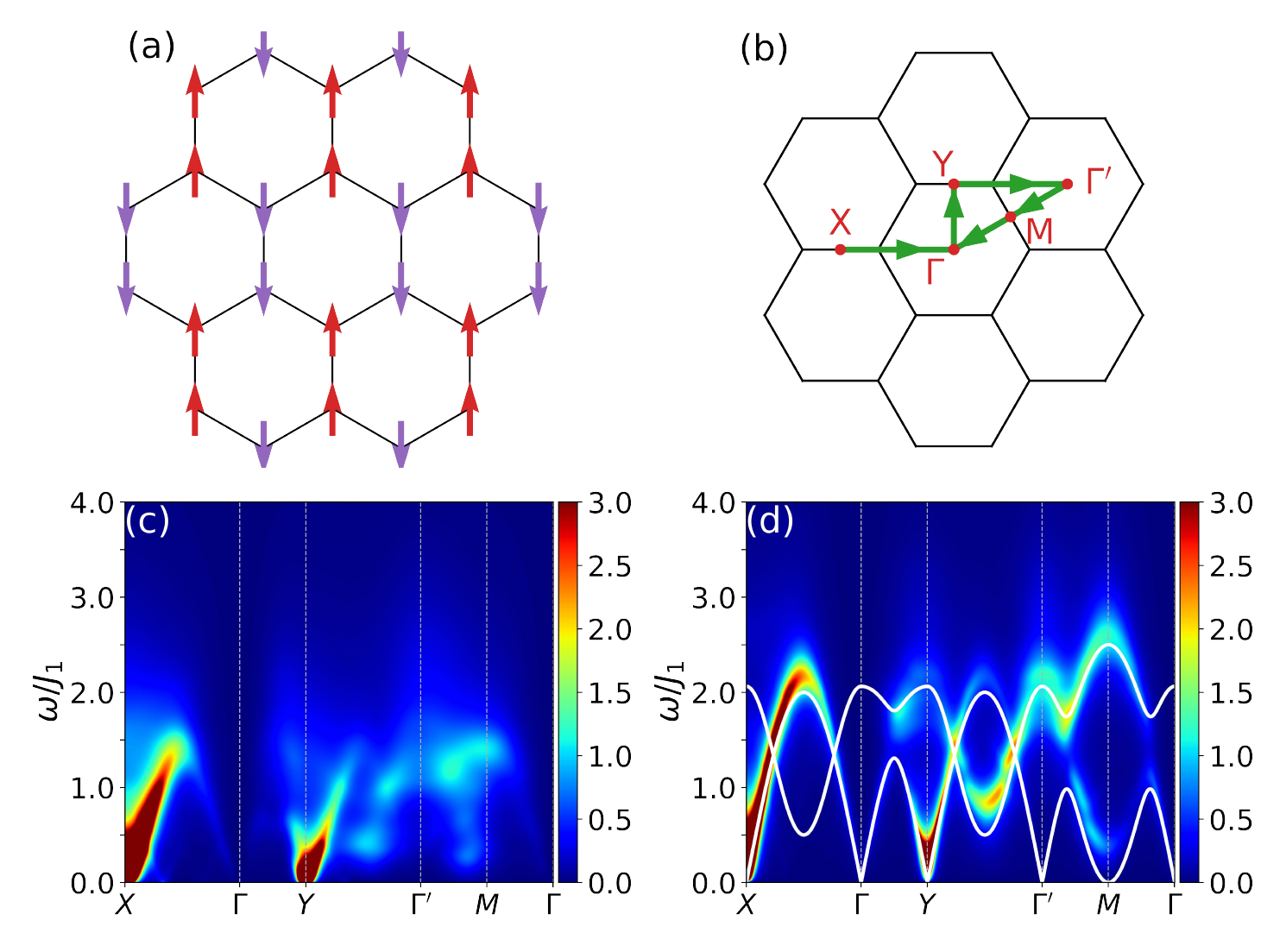}
  \caption{\label{fig:stripe-order}(Color online) (a) Illustration of the pattern of the stripe magnetic order. (b) Path for the illustration of the spectrum in the stipe phase. (c) and (d) Dynamical structure factors $S^{+-}(\bm{k},\omega)$ for $J_{2}/J_{1}=0.5$, $J_{3}/J_{1}=0$ and $J_{2}/J_{1}=0.3$, $J_{3}/J_{1}=-0.5$, respectively. The white line in (d) is the LSW dispersion with the $1/S$ correction.}
\end{figure}
When $J_{2}>0.48J_{1}$, we find that the system enters into another magnetic phase with a stripe order as shown in Fig.~\ref{fig:stripe-order}(a), which is consistent with the results obtained by the VMC\cite{PhysRevB.89.094413}, the modified spin wave theory \cite{2016Quantum} and coupled-cluster method \cite{PhysRevB.89.220408}. In the classical limit, there are infinite number of degenerate magnetic structures, whose wave vectors form a closed contour in the BZ \cite{PhysRevB.81.214419}, and the wave vector for the stripe order is not on this contour necessarily for all $J_{2}$. This is a further evidence that the Heisenberg model on the honeycomb lattice has strong quantum fluctuations, which substantially change the classical scenario and stabilize the stripe order for large $J_{2}$. In order to reflect the rotational symmetry breaking of the stripe order, we choose a different path in BZ [see Fig.~\ref{fig:stripe-order}(b)] to present the excitation spectrum. The result for $J_{2}=0.5J_{1}$ is shown in Fig.~\ref{fig:stripe-order}(c). One can see that the Goldstone modes appear at $X$ and $Y$, which is a key feature of the stripe order. Since the stripe order is not necessarily the classical ground state for the $J_{1}$-$J_{2}$ model, the LSW theory can not be applied to the stripe order. In order to reveal the effects of quantum fluctuations on the excitation spectrum by comparing the results obtained by the CPT and LSW methods, we introduce the third-nearest-neighbor Heisenberg exchange $J_{3}$ into the model, which can stabilize the stripe order at the classical level. Figure~\ref{fig:stripe-order}(d) shows the CPT spectrum together with the LSW dispersion for $J_{2}=0.3J_{1}$ and $J_{3}=-0.5J_{1}$, for which the classical ground state has a stabilized stripe magnetic order. Although the dispersions based on the two theoretical methods are consistent at most momentum point, a remarkable difference is that the CPT spectrum is gapped at the $M$ point while the LSW spectrum is gapless, which is a strong manifest of the effects of quantum fluctuations. We note that this gap at the $M$ point also exists in the spectrum for $J_{3}=0$ as shown in Fig.~\ref{fig:stripe-order}(c). Thus, the quantum fluctuations not only selects the stripe order as the ground state, but also leads to a gap in the excitation spectrum at the $M$ point.

\subsection{\label{sec:E} Comparison of CPT results in N\'{e}el phase to experiments}
In order to directly compare with the experimental results, we also exhibit the spectrum for $J_{2}=0.1J_{1}$ in Fig.~\ref{fig:neel-order}(e) along the same path [see Fig.~\ref{fig:neel-order}(d)] as that used in the experiment \cite{2021Van}. We find that the result for $J_{2}=0.1J_{1}$ reproduces the following features observed in the experiment~\cite{2021Van}. Namely, (i) besides the well-defined low-energy magnon excitations, there is an obvious dome-shaped continuum up to twice the energy of the magnon band top and centering around the $\Gamma^{\prime}$ point; (ii) the continuum disappears in a large region centering around $\Gamma$ (the center of the first BZ), as can also be seen in Fig.~\ref{fig:neel-order}(c); (iii) around the $K$ ($K^{\prime}$) point, the spectral weights are suppressed and no well-defined magnon mode can be identified, so the spectrum there exhibits an additional continuum. At the meantime, the characteristic that the additional strong excitation continua around the $K$ ($K^{\prime}$) point and along the $K-M^{\prime}$ ($K^{\prime}-Y^{\prime}$) line coexisting with the low-energy magnon excitations which disperse from the BZ center for $J_{2}=0.1J_{1}$ are also consistent with the neutron scattering results on its sister compound YbBr$_{3}$ \cite{npjqm.5.85}. According to the discussion of the possible origin of the additional strong continua presented in Subsection C, we propose that the continua observed at the $K$ point of the BZ in the INS experiments on YbCl$_{3}$ \cite{2021Van} and YbBr$_{3}$ \cite{npjqm.5.85} may be mainly due to the fractionalization of the $S=1$ spin excitations into deconfined spinons.

On the other hand, we note that an observable feature with high intensity along the upper boundary of the dome-shaped continua has also been identified in YbCl$_{3}$ \cite{2021Van}, which is absent in our calculation. As shown in Ref.\onlinecite{2021Van}, the dome-shaped continuum can be accounted for with the LSW theory by including both the transverse and longitudinal spin scattering channels and the high intensity feature within the continuum is ascribed to be a Van Hove singularity in a two-magnon continuum which occurs only in the longitudinal scattering channels. Here in our theoretical calculations, the longitudinal spin fluctuations are not included. Our qualitative reproduction of the dome-shaped continuum only in the transverse scattering channel would suggest that the continuum observed in the INS measurement in YbCl$_{3}$ may come from not only the longitudinal spin excitations, but also the transversal spin excitations. This is different from the LSW scenario in Ref.\onlinecite{2021Van}, where the continuum comes only from the longitudinal spin excitations and the transversal spin excitations contribute only the sharp spin-wave dispersion.

\section{\label{sec:SectionVI}Summary}

In summary, we have investigated the spin dynamics of the antiferromagnetic $J_{1}$-$J_{2}$ Heisenberg model on the honeycomb lattice making use of the spin cluster perturbation theory. We obtain the excitation spectra of all four possible phases for different $J_{2}$: a N\'{e}el phase, a plaquette valence-bond-solid phase, a dimer valence-bond-solid phase and a stripe antiferromagnetic phase. In the unfrustrated Heisenberg model with $J_2=0$, we have already found clear deviation from the magnon dispersion obtained by the linear-spin-wave calculation along the boundary of the Brillouin zone, namely the band is relatively flat. The excitation spectrum for the case of $J_{2}=0.1J_{1}$ is consistent with the recent inelastic neutron scattering measurement on YbCl$_{3}$ and YbBr$_{3}$, including the dispersion, the dome-shaped continuum around the center of the second Brillouin zone and the additional strong continuum close to the corner of the Brillouin zone. In the valence-bond-solid phases, the spectra show a clear gap and a broad continuum, and different triplon excitation dispersions at lowest energies corresponding to plaquette and dimer valence-bond-solid phases, respectively. This broad continuum is shown to coming from the nearly-deconfined spinon excitations. Moreover, the continuous spectrum in the plaquette valence-bond-solid phase evolves into the additional strong continuum near the corner of the Brillouin zone in the N\'{e}el phase. Thus, our results reveal that the additional continuum in the N\'{e}el phase have indispensable contributions from the deconfinement of fractionalized spin-$1/2$ spinons. Finally, in the stripe state, the deviation of the spectrum from the result in the linear-spin-wave calculation occurs at the $M$ point, at which the spectrum is gapped in contrast to the gapless spectrum in the latter, which is believed to be a strong effect of quantum fluctuations.

\begin{acknowledgments}
This work was supported by National Key Projects for Research and Development of China (Grant No. 2021YFA1400400) and the National Natural Science Foundation of China (No. 92165205 and No. 12074175).
\end{acknowledgments}

\appendix
\section{CPT spectra for the N\'{e}el and VBS phases with logarithmic colorbar}
\label{appendix-a}

\begin{figure}
  \includegraphics[width=\columnwidth]{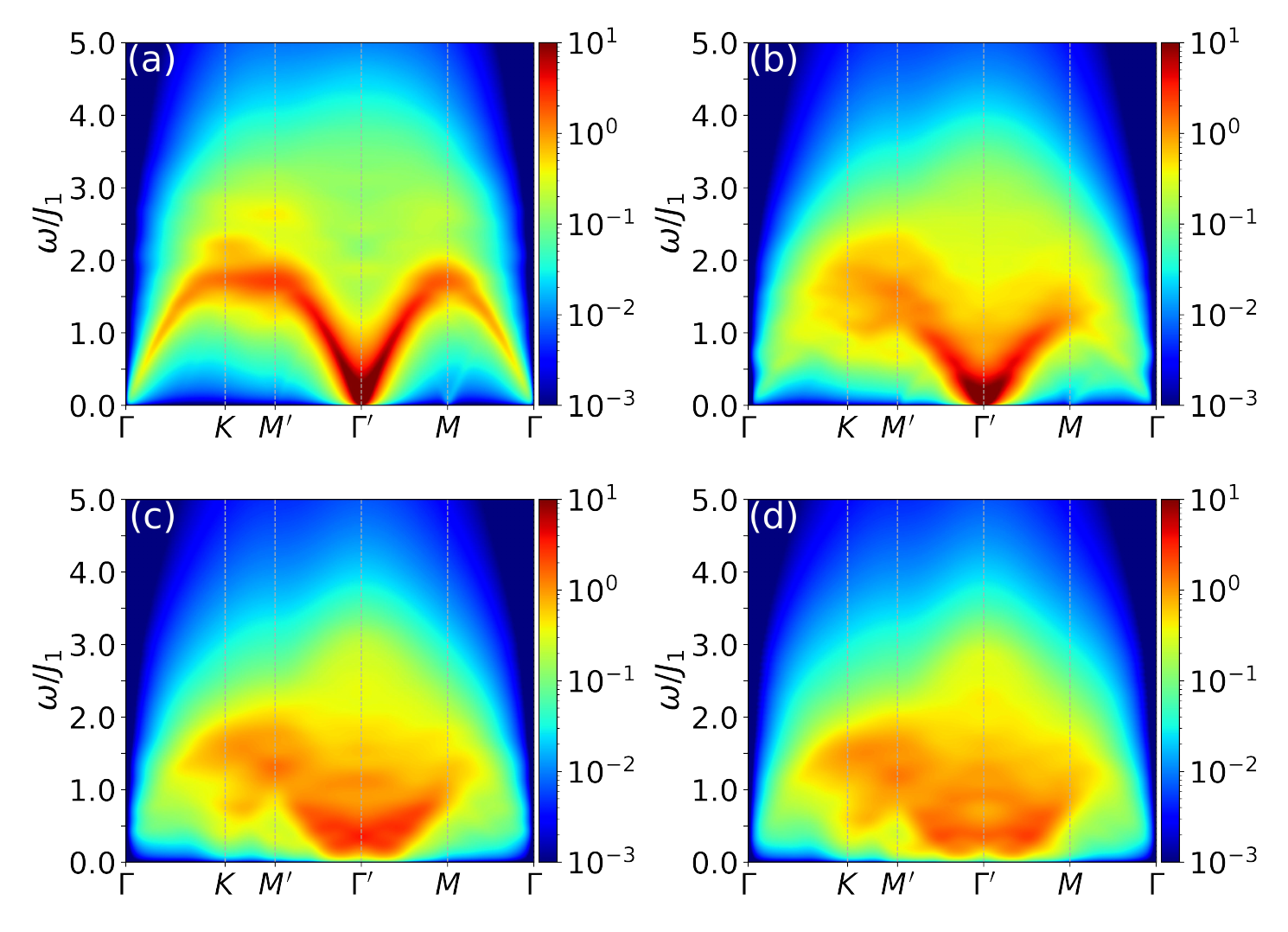}
  \caption{\label{fig:appendix}(Color online) Dynamical structure factor $S^{+-}(\bm{k},\omega)$ with the logarithmic colorbar for (a) $J_{2}/J_{1}=0$, (b) $J_{2}/J_{1}=0.15$, (c) $J_{2}/J_{1}=0.3$ and (d) $J_{2}/J_{1}=0.37$.}
\end{figure}

In the main text, the intensities of our CPT spectra are exhibited with a linear colorbar, which  is also the usual scheme used in experimental works. Nevertheless, to reveal the detailed features of the spectra, the logarithmic colorbar is an alternative~\cite{ferrari2020dynamical}. In Fig.~\ref{fig:appendix}, we display the CPT spectra with the same logarithmic colorbar as that in the VMC study~\cite{ferrari2020dynamical}, which is helpful to show more clearly the relatively weak details and make a direct comparison between the results obtained by the two methods.  We can find that the overall characteristics of the CPT spectra are consistent with those given by the VMC calculation~\cite{ferrari2020dynamical}. As a comparison to our CPT results with a linear colorbar shown in Fig.~\ref{fig:neel-order} in the main text, we find that, i) both the dome-shaped continuum centered on the $\Gamma'$ point and the additional continuum around the $K$ point show up more clearly, ii) the Goldstone mode at the $\Gamma$ point now can be identified directly in the figure.

\bibliography{ref}

\begin{thebibliography}{61}%
\makeatletter
\providecommand \@ifxundefined [1]{%
 \@ifx{#1\undefined}
}%
\providecommand \@ifnum [1]{%
 \ifnum #1\expandafter \@firstoftwo
 \else \expandafter \@secondoftwo
 \fi
}%
\providecommand \@ifx [1]{%
 \ifx #1\expandafter \@firstoftwo
 \else \expandafter \@secondoftwo
 \fi
}%
\providecommand \natexlab [1]{#1}%
\providecommand \enquote  [1]{``#1''}%
\providecommand \bibnamefont  [1]{#1}%
\providecommand \bibfnamefont [1]{#1}%
\providecommand \citenamefont [1]{#1}%
\providecommand \href@noop [0]{\@secondoftwo}%
\providecommand \href [0]{\begingroup \@sanitize@url \@href}%
\providecommand \@href[1]{\@@startlink{#1}\@@href}%
\providecommand \@@href[1]{\endgroup#1\@@endlink}%
\providecommand \@sanitize@url [0]{\catcode `\\12\catcode `\$12\catcode
  `\&12\catcode `\#12\catcode `\^12\catcode `\_12\catcode `\%12\relax}%
\providecommand \@@startlink[1]{}%
\providecommand \@@endlink[0]{}%
\providecommand \url  [0]{\begingroup\@sanitize@url \@url }%
\providecommand \@url [1]{\endgroup\@href {#1}{\urlprefix }}%
\providecommand \urlprefix  [0]{URL }%
\providecommand \Eprint [0]{\href }%
\providecommand \doibase [0]{http://dx.doi.org/}%
\providecommand \selectlanguage [0]{\@gobble}%
\providecommand \bibinfo  [0]{\@secondoftwo}%
\providecommand \bibfield  [0]{\@secondoftwo}%
\providecommand \translation [1]{[#1]}%
\providecommand \BibitemOpen [0]{}%
\providecommand \bibitemStop [0]{}%
\providecommand \bibitemNoStop [0]{.\EOS\space}%
\providecommand \EOS [0]{\spacefactor3000\relax}%
\providecommand \BibitemShut  [1]{\csname bibitem#1\endcsname}%
\let\auto@bib@innerbib\@empty
%</preamble>
\bibitem [{\citenamefont {Chakravarty}\ \emph {et~al.}(1989)\citenamefont
  {Chakravarty}, \citenamefont {Halperin},\ and\ \citenamefont
  {Nelson}}]{PhysRevB.39.2344}%
  \BibitemOpen
  \bibfield  {author} {\bibinfo {author} {\bibfnamefont {S.}~\bibnamefont
  {Chakravarty}}, \bibinfo {author} {\bibfnamefont {B.~I.}\ \bibnamefont
  {Halperin}}, \ and\ \bibinfo {author} {\bibfnamefont {D.~R.}\ \bibnamefont
  {Nelson}},\ }\href {\doibase 10.1103/PhysRevB.39.2344} {\bibfield  {journal}
  {\bibinfo  {journal} {Phys. Rev. B}\ }\textbf {\bibinfo {volume} {39}},\
  \bibinfo {pages} {2344} (\bibinfo {year} {1989})}\BibitemShut {NoStop}%
\bibitem [{\citenamefont {Manousakis}(1991)}]{RevModPhys.63.1}%
  \BibitemOpen
  \bibfield  {author} {\bibinfo {author} {\bibfnamefont {E.}~\bibnamefont
  {Manousakis}},\ }\href {\doibase 10.1103/RevModPhys.63.1} {\bibfield
  {journal} {\bibinfo  {journal} {Rev. Mod. Phys.}\ }\textbf {\bibinfo {volume}
  {63}},\ \bibinfo {pages} {1} (\bibinfo {year} {1991})}\BibitemShut {NoStop}%
\bibitem [{\citenamefont {Sachdev}(2008)}]{sachdev2008}%
  \BibitemOpen
  \bibfield  {author} {\bibinfo {author} {\bibfnamefont {S.}~\bibnamefont
  {Sachdev}},\ }\href {\doibase 10.1038/nphys894} {\bibfield  {journal}
  {\bibinfo  {journal} {Nat. Phys.}\ }\textbf {\bibinfo {volume} {4}},\
  \bibinfo {pages} {173} (\bibinfo {year} {2008})}\BibitemShut {NoStop}%
\bibitem [{\citenamefont {Savary}\ and\ \citenamefont
  {Balents}(2016)}]{Savary_2016}%
  \BibitemOpen
  \bibfield  {author} {\bibinfo {author} {\bibfnamefont {L.}~\bibnamefont
  {Savary}}\ and\ \bibinfo {author} {\bibfnamefont {L.}~\bibnamefont
  {Balents}},\ }\href {\doibase 10.1088/0034-4885/80/1/016502} {\bibfield
  {journal} {\bibinfo  {journal} {Reports on Progress in Physics}\ }\textbf
  {\bibinfo {volume} {80}},\ \bibinfo {pages} {016502} (\bibinfo {year}
  {2016})}\BibitemShut {NoStop}%
\bibitem [{\citenamefont {Broholm}\ \emph {et~al.}(2020)\citenamefont
  {Broholm}, \citenamefont {Cava}, \citenamefont {Kivelson}, \citenamefont
  {Nocera}, \citenamefont {Norman},\ and\ \citenamefont
  {Senthil}}]{Broholmeaay0668}%
  \BibitemOpen
  \bibfield  {author} {\bibinfo {author} {\bibfnamefont {C.}~\bibnamefont
  {Broholm}}, \bibinfo {author} {\bibfnamefont {R.~J.}\ \bibnamefont {Cava}},
  \bibinfo {author} {\bibfnamefont {S.~A.}\ \bibnamefont {Kivelson}}, \bibinfo
  {author} {\bibfnamefont {D.~G.}\ \bibnamefont {Nocera}}, \bibinfo {author}
  {\bibfnamefont {M.~R.}\ \bibnamefont {Norman}}, \ and\ \bibinfo {author}
  {\bibfnamefont {T.}~\bibnamefont {Senthil}},\ }\href
  {https://www.science.org/doi/10.1126/science.aay0668} {\bibfield  {journal}
  {\bibinfo  {journal} {Science}\ }\textbf {\bibinfo {volume} {367}} (\bibinfo
  {year} {2020})}\BibitemShut {NoStop}%
\bibitem [{\citenamefont {Zhou}\ \emph {et~al.}(2017)\citenamefont {Zhou},
  \citenamefont {Kanoda},\ and\ \citenamefont {Ng}}]{RevModPhys.89.025003}%
  \BibitemOpen
  \bibfield  {author} {\bibinfo {author} {\bibfnamefont {Y.}~\bibnamefont
  {Zhou}}, \bibinfo {author} {\bibfnamefont {K.}~\bibnamefont {Kanoda}}, \ and\
  \bibinfo {author} {\bibfnamefont {T.-K.}\ \bibnamefont {Ng}},\ }\href
  {\doibase 10.1103/RevModPhys.89.025003} {\bibfield  {journal} {\bibinfo
  {journal} {Rev. Mod. Phys.}\ }\textbf {\bibinfo {volume} {89}},\ \bibinfo
  {pages} {025003} (\bibinfo {year} {2017})}\BibitemShut {NoStop}%
\bibitem [{\citenamefont {Capriotti}\ \emph {et~al.}(2001)\citenamefont
  {Capriotti}, \citenamefont {Becca}, \citenamefont {Parola},\ and\
  \citenamefont {Sorella}}]{PhysRevLett.87.097201}%
  \BibitemOpen
  \bibfield  {author} {\bibinfo {author} {\bibfnamefont {L.}~\bibnamefont
  {Capriotti}}, \bibinfo {author} {\bibfnamefont {F.}~\bibnamefont {Becca}},
  \bibinfo {author} {\bibfnamefont {A.}~\bibnamefont {Parola}}, \ and\ \bibinfo
  {author} {\bibfnamefont {S.}~\bibnamefont {Sorella}},\ }\href {\doibase
  10.1103/PhysRevLett.87.097201} {\bibfield  {journal} {\bibinfo  {journal}
  {Phys. Rev. Lett.}\ }\textbf {\bibinfo {volume} {87}},\ \bibinfo {pages}
  {097201} (\bibinfo {year} {2001})}\BibitemShut {NoStop}%
\bibitem [{\citenamefont {Sirker}\ \emph {et~al.}(2006)\citenamefont {Sirker},
  \citenamefont {Weihong}, \citenamefont {Sushkov},\ and\ \citenamefont
  {Oitmaa}}]{PhysRevB.73.184420}%
  \BibitemOpen
  \bibfield  {author} {\bibinfo {author} {\bibfnamefont {J.}~\bibnamefont
  {Sirker}}, \bibinfo {author} {\bibfnamefont {Z.}~\bibnamefont {Weihong}},
  \bibinfo {author} {\bibfnamefont {O.~P.}\ \bibnamefont {Sushkov}}, \ and\
  \bibinfo {author} {\bibfnamefont {J.}~\bibnamefont {Oitmaa}},\ }\href
  {\doibase 10.1103/PhysRevB.73.184420} {\bibfield  {journal} {\bibinfo
  {journal} {Phys. Rev. B}\ }\textbf {\bibinfo {volume} {73}},\ \bibinfo
  {pages} {184420} (\bibinfo {year} {2006})}\BibitemShut {NoStop}%
\bibitem [{\citenamefont {Jiang}\ \emph {et~al.}(2012)\citenamefont {Jiang},
  \citenamefont {Yao},\ and\ \citenamefont {Balents}}]{PhysRevB.86.024424}%
  \BibitemOpen
  \bibfield  {author} {\bibinfo {author} {\bibfnamefont {H.-C.}\ \bibnamefont
  {Jiang}}, \bibinfo {author} {\bibfnamefont {H.}~\bibnamefont {Yao}}, \ and\
  \bibinfo {author} {\bibfnamefont {L.}~\bibnamefont {Balents}},\ }\href
  {\doibase 10.1103/PhysRevB.86.024424} {\bibfield  {journal} {\bibinfo
  {journal} {Phys. Rev. B}\ }\textbf {\bibinfo {volume} {86}},\ \bibinfo
  {pages} {024424} (\bibinfo {year} {2012})}\BibitemShut {NoStop}%
\bibitem [{\citenamefont {Wang}\ \emph {et~al.}(2013)\citenamefont {Wang},
  \citenamefont {Poilblanc}, \citenamefont {Gu}, \citenamefont {Wen},\ and\
  \citenamefont {Verstraete}}]{PhysRevLett.111.037202}%
  \BibitemOpen
  \bibfield  {author} {\bibinfo {author} {\bibfnamefont {L.}~\bibnamefont
  {Wang}}, \bibinfo {author} {\bibfnamefont {D.}~\bibnamefont {Poilblanc}},
  \bibinfo {author} {\bibfnamefont {Z.-C.}\ \bibnamefont {Gu}}, \bibinfo
  {author} {\bibfnamefont {X.-G.}\ \bibnamefont {Wen}}, \ and\ \bibinfo
  {author} {\bibfnamefont {F.}~\bibnamefont {Verstraete}},\ }\href {\doibase
  10.1103/PhysRevLett.111.037202} {\bibfield  {journal} {\bibinfo  {journal}
  {Phys. Rev. Lett.}\ }\textbf {\bibinfo {volume} {111}},\ \bibinfo {pages}
  {037202} (\bibinfo {year} {2013})}\BibitemShut {NoStop}%
\bibitem [{\citenamefont {Gong}\ \emph {et~al.}(2014)\citenamefont {Gong},
  \citenamefont {Zhu}, \citenamefont {Sheng}, \citenamefont {Motrunich},\ and\
  \citenamefont {Fisher}}]{PhysRevLett.113.027201}%
  \BibitemOpen
  \bibfield  {author} {\bibinfo {author} {\bibfnamefont {S.-S.}\ \bibnamefont
  {Gong}}, \bibinfo {author} {\bibfnamefont {W.}~\bibnamefont {Zhu}}, \bibinfo
  {author} {\bibfnamefont {D.~N.}\ \bibnamefont {Sheng}}, \bibinfo {author}
  {\bibfnamefont {O.~I.}\ \bibnamefont {Motrunich}}, \ and\ \bibinfo {author}
  {\bibfnamefont {M.~P.~A.}\ \bibnamefont {Fisher}},\ }\href {\doibase
  10.1103/PhysRevLett.113.027201} {\bibfield  {journal} {\bibinfo  {journal}
  {Phys. Rev. Lett.}\ }\textbf {\bibinfo {volume} {113}},\ \bibinfo {pages}
  {027201} (\bibinfo {year} {2014})}\BibitemShut {NoStop}%
\bibitem [{\citenamefont {Poilblanc}\ and\ \citenamefont
  {Mambrini}(2017)}]{PhysRevB.96.014414}%
  \BibitemOpen
  \bibfield  {author} {\bibinfo {author} {\bibfnamefont {D.}~\bibnamefont
  {Poilblanc}}\ and\ \bibinfo {author} {\bibfnamefont {M.}~\bibnamefont
  {Mambrini}},\ }\href {\doibase 10.1103/PhysRevB.96.014414} {\bibfield
  {journal} {\bibinfo  {journal} {Phys. Rev. B}\ }\textbf {\bibinfo {volume}
  {96}},\ \bibinfo {pages} {014414} (\bibinfo {year} {2017})}\BibitemShut
  {NoStop}%
\bibitem [{\citenamefont {Haghshenas}\ and\ \citenamefont
  {Sheng}(2018)}]{PhysRevB.97.174408}%
  \BibitemOpen
  \bibfield  {author} {\bibinfo {author} {\bibfnamefont {R.}~\bibnamefont
  {Haghshenas}}\ and\ \bibinfo {author} {\bibfnamefont {D.~N.}\ \bibnamefont
  {Sheng}},\ }\href {\doibase 10.1103/PhysRevB.97.174408} {\bibfield  {journal}
  {\bibinfo  {journal} {Phys. Rev. B}\ }\textbf {\bibinfo {volume} {97}},\
  \bibinfo {pages} {174408} (\bibinfo {year} {2018})}\BibitemShut {NoStop}%
\bibitem [{\citenamefont {Wang}\ and\ \citenamefont
  {Sandvik}(2018)}]{PhysRevLett.121.107202}%
  \BibitemOpen
  \bibfield  {author} {\bibinfo {author} {\bibfnamefont {L.}~\bibnamefont
  {Wang}}\ and\ \bibinfo {author} {\bibfnamefont {A.~W.}\ \bibnamefont
  {Sandvik}},\ }\href {\doibase 10.1103/PhysRevLett.121.107202} {\bibfield
  {journal} {\bibinfo  {journal} {Phys. Rev. Lett.}\ }\textbf {\bibinfo
  {volume} {121}},\ \bibinfo {pages} {107202} (\bibinfo {year}
  {2018})}\BibitemShut {NoStop}%
\bibitem [{\citenamefont {Nomura}\ and\ \citenamefont
  {Imada}(2021)}]{PhysRevX.11.031034}%
  \BibitemOpen
  \bibfield  {author} {\bibinfo {author} {\bibfnamefont {Y.}~\bibnamefont
  {Nomura}}\ and\ \bibinfo {author} {\bibfnamefont {M.}~\bibnamefont {Imada}},\
  }\href {\doibase 10.1103/PhysRevX.11.031034} {\bibfield  {journal} {\bibinfo
  {journal} {Phys. Rev. X}\ }\textbf {\bibinfo {volume} {11}},\ \bibinfo
  {pages} {031034} (\bibinfo {year} {2021})}\BibitemShut {NoStop}%
\bibitem [{\citenamefont {Piazza}\ \emph {et~al.}(2015)\citenamefont {Piazza},
  \citenamefont {Mourigal}, \citenamefont {Christensen}, \citenamefont
  {Nilsen}, \citenamefont {Tregenna-Piggott}, \citenamefont {Perring},
  \citenamefont {Enderle}, \citenamefont {McMorrow}, \citenamefont {Ivanov},\
  and\ \citenamefont {Rønnow}}]{Naturephys2015}%
  \BibitemOpen
  \bibfield  {author} {\bibinfo {author} {\bibfnamefont {D.}~\bibnamefont
  {Piazza}}, \bibinfo {author} {\bibfnamefont {B.}~\bibnamefont {Mourigal}},
  \bibinfo {author} {\bibfnamefont {M.}~\bibnamefont {Christensen}}, \bibinfo
  {author} {\bibfnamefont {G.~J.}\ \bibnamefont {Nilsen}}, \bibinfo {author}
  {\bibfnamefont {P.}~\bibnamefont {Tregenna-Piggott}}, \bibinfo {author}
  {\bibfnamefont {T.~G.}\ \bibnamefont {Perring}}, \bibinfo {author}
  {\bibfnamefont {M.}~\bibnamefont {Enderle}}, \bibinfo {author} {\bibfnamefont
  {D.~F.}\ \bibnamefont {McMorrow}}, \bibinfo {author} {\bibfnamefont {D.~A.}\
  \bibnamefont {Ivanov}}, \ and\ \bibinfo {author} {\bibfnamefont {H.~M.}\
  \bibnamefont {Rønnow}},\ }\href {\doibase doi.org/10.1038/nphys3172}
  {\bibfield  {journal} {\bibinfo  {journal} {Nature Phys}\ }\textbf {\bibinfo
  {volume} {11}},\ \bibinfo {pages} {62} (\bibinfo {year} {2015})}\BibitemShut
  {NoStop}%
\bibitem [{\citenamefont {Headings}\ \emph {et~al.}(2010)\citenamefont
  {Headings}, \citenamefont {Hayden}, \citenamefont {Coldea},\ and\
  \citenamefont {Perring}}]{PhysRevLett.105.247001}%
  \BibitemOpen
  \bibfield  {author} {\bibinfo {author} {\bibfnamefont {N.~S.}\ \bibnamefont
  {Headings}}, \bibinfo {author} {\bibfnamefont {S.~M.}\ \bibnamefont
  {Hayden}}, \bibinfo {author} {\bibfnamefont {R.}~\bibnamefont {Coldea}}, \
  and\ \bibinfo {author} {\bibfnamefont {T.~G.}\ \bibnamefont {Perring}},\
  }\href {\doibase 10.1103/PhysRevLett.105.247001} {\bibfield  {journal}
  {\bibinfo  {journal} {Phys. Rev. Lett.}\ }\textbf {\bibinfo {volume} {105}},\
  \bibinfo {pages} {247001} (\bibinfo {year} {2010})}\BibitemShut {NoStop}%
\bibitem [{\citenamefont {Shao}\ \emph {et~al.}(2017)\citenamefont {Shao},
  \citenamefont {Qin}, \citenamefont {Capponi}, \citenamefont {Chesi},
  \citenamefont {Meng},\ and\ \citenamefont {Sandvik}}]{PhysRevX.7.041072}%
  \BibitemOpen
  \bibfield  {author} {\bibinfo {author} {\bibfnamefont {H.}~\bibnamefont
  {Shao}}, \bibinfo {author} {\bibfnamefont {Y.~Q.}\ \bibnamefont {Qin}},
  \bibinfo {author} {\bibfnamefont {S.}~\bibnamefont {Capponi}}, \bibinfo
  {author} {\bibfnamefont {S.}~\bibnamefont {Chesi}}, \bibinfo {author}
  {\bibfnamefont {Z.~Y.}\ \bibnamefont {Meng}}, \ and\ \bibinfo {author}
  {\bibfnamefont {A.~W.}\ \bibnamefont {Sandvik}},\ }\href {\doibase
  10.1103/PhysRevX.7.041072} {\bibfield  {journal} {\bibinfo  {journal} {Phys.
  Rev. X}\ }\textbf {\bibinfo {volume} {7}},\ \bibinfo {pages} {041072}
  (\bibinfo {year} {2017})}\BibitemShut {NoStop}%
\bibitem [{\citenamefont {Singh}\ and\ \citenamefont
  {Gelfand}(1995)}]{PhysRevB.52.R15695}%
  \BibitemOpen
  \bibfield  {author} {\bibinfo {author} {\bibfnamefont {R.~R.~P.}\
  \bibnamefont {Singh}}\ and\ \bibinfo {author} {\bibfnamefont {M.~P.}\
  \bibnamefont {Gelfand}},\ }\href {\doibase 10.1103/PhysRevB.52.R15695}
  {\bibfield  {journal} {\bibinfo  {journal} {Phys. Rev. B}\ }\textbf {\bibinfo
  {volume} {52}},\ \bibinfo {pages} {R15695} (\bibinfo {year}
  {1995})}\BibitemShut {NoStop}%
\bibitem [{\citenamefont {Sandvik}\ and\ \citenamefont
  {Singh}(2001)}]{PhysRevLett.86.528}%
  \BibitemOpen
  \bibfield  {author} {\bibinfo {author} {\bibfnamefont {A.~W.}\ \bibnamefont
  {Sandvik}}\ and\ \bibinfo {author} {\bibfnamefont {R.~R.~P.}\ \bibnamefont
  {Singh}},\ }\href {\doibase 10.1103/PhysRevLett.86.528} {\bibfield  {journal}
  {\bibinfo  {journal} {Phys. Rev. Lett.}\ }\textbf {\bibinfo {volume} {86}},\
  \bibinfo {pages} {528} (\bibinfo {year} {2001})}\BibitemShut {NoStop}%
\bibitem [{\citenamefont {Powalski}\ \emph {et~al.}(2018)\citenamefont
  {Powalski}, \citenamefont {Schmidt},\ and\ \citenamefont
  {Uhrig}}]{SciPostPhys.4.1.001}%
  \BibitemOpen
  \bibfield  {author} {\bibinfo {author} {\bibfnamefont {M.}~\bibnamefont
  {Powalski}}, \bibinfo {author} {\bibfnamefont {K.~P.}\ \bibnamefont
  {Schmidt}}, \ and\ \bibinfo {author} {\bibfnamefont {G.~S.}\ \bibnamefont
  {Uhrig}},\ }\href {\doibase 10.21468/SciPostPhys.4.1.001} {\bibfield
  {journal} {\bibinfo  {journal} {SciPost Phys.}\ }\textbf {\bibinfo {volume}
  {4}},\ \bibinfo {pages} {001} (\bibinfo {year} {2018})}\BibitemShut {NoStop}%
\bibitem [{\citenamefont {Ferrari}\ and\ \citenamefont
  {Becca}(2018)}]{PhysRevB.98.100405}%
  \BibitemOpen
  \bibfield  {author} {\bibinfo {author} {\bibfnamefont {F.}~\bibnamefont
  {Ferrari}}\ and\ \bibinfo {author} {\bibfnamefont {F.}~\bibnamefont
  {Becca}},\ }\href {\doibase 10.1103/PhysRevB.98.100405} {\bibfield  {journal}
  {\bibinfo  {journal} {Phys. Rev. B}\ }\textbf {\bibinfo {volume} {98}},\
  \bibinfo {pages} {100405} (\bibinfo {year} {2018})}\BibitemShut {NoStop}%
\bibitem [{\citenamefont {Yu}\ \emph {et~al.}(2018)\citenamefont {Yu},
  \citenamefont {Wang}, \citenamefont {Dong}, \citenamefont {Yao},\ and\
  \citenamefont {Li}}]{PhysRevB.98.134410}%
  \BibitemOpen
  \bibfield  {author} {\bibinfo {author} {\bibfnamefont {S.-L.}\ \bibnamefont
  {Yu}}, \bibinfo {author} {\bibfnamefont {W.}~\bibnamefont {Wang}}, \bibinfo
  {author} {\bibfnamefont {Z.-Y.}\ \bibnamefont {Dong}}, \bibinfo {author}
  {\bibfnamefont {Z.-J.}\ \bibnamefont {Yao}}, \ and\ \bibinfo {author}
  {\bibfnamefont {J.-X.}\ \bibnamefont {Li}},\ }\href {\doibase
  10.1103/PhysRevB.98.134410} {\bibfield  {journal} {\bibinfo  {journal} {Phys.
  Rev. B}\ }\textbf {\bibinfo {volume} {98}},\ \bibinfo {pages} {134410}
  (\bibinfo {year} {2018})}\BibitemShut {NoStop}%
\bibitem [{\citenamefont {Yamamoto}\ and\ \citenamefont
  {Noriki}(2019)}]{PhysRevB.99.094412}%
  \BibitemOpen
  \bibfield  {author} {\bibinfo {author} {\bibfnamefont {S.}~\bibnamefont
  {Yamamoto}}\ and\ \bibinfo {author} {\bibfnamefont {Y.}~\bibnamefont
  {Noriki}},\ }\href {\doibase 10.1103/PhysRevB.99.094412} {\bibfield
  {journal} {\bibinfo  {journal} {Phys. Rev. B}\ }\textbf {\bibinfo {volume}
  {99}},\ \bibinfo {pages} {094412} (\bibinfo {year} {2019})}\BibitemShut
  {NoStop}%
\bibitem [{\citenamefont {Zheng}\ \emph {et~al.}(2005)\citenamefont {Zheng},
  \citenamefont {Oitmaa},\ and\ \citenamefont {Hamer}}]{PhysRevB.71.184440}%
  \BibitemOpen
  \bibfield  {author} {\bibinfo {author} {\bibfnamefont {W.}~\bibnamefont
  {Zheng}}, \bibinfo {author} {\bibfnamefont {J.}~\bibnamefont {Oitmaa}}, \
  and\ \bibinfo {author} {\bibfnamefont {C.~J.}\ \bibnamefont {Hamer}},\ }\href
  {\doibase 10.1103/PhysRevB.71.184440} {\bibfield  {journal} {\bibinfo
  {journal} {Phys. Rev. B}\ }\textbf {\bibinfo {volume} {71}},\ \bibinfo
  {pages} {184440} (\bibinfo {year} {2005})}\BibitemShut {NoStop}%
\bibitem [{\citenamefont {Coldea}\ \emph {et~al.}(2001)\citenamefont {Coldea},
  \citenamefont {Hayden}, \citenamefont {Aeppli}, \citenamefont {Perring},
  \citenamefont {Frost}, \citenamefont {Mason}, \citenamefont {Cheong},\ and\
  \citenamefont {Fisk}}]{PhysRevLett.86.5377}%
  \BibitemOpen
  \bibfield  {author} {\bibinfo {author} {\bibfnamefont {R.}~\bibnamefont
  {Coldea}}, \bibinfo {author} {\bibfnamefont {S.~M.}\ \bibnamefont {Hayden}},
  \bibinfo {author} {\bibfnamefont {G.}~\bibnamefont {Aeppli}}, \bibinfo
  {author} {\bibfnamefont {T.~G.}\ \bibnamefont {Perring}}, \bibinfo {author}
  {\bibfnamefont {C.~D.}\ \bibnamefont {Frost}}, \bibinfo {author}
  {\bibfnamefont {T.~E.}\ \bibnamefont {Mason}}, \bibinfo {author}
  {\bibfnamefont {S.-W.}\ \bibnamefont {Cheong}}, \ and\ \bibinfo {author}
  {\bibfnamefont {Z.}~\bibnamefont {Fisk}},\ }\href {\doibase
  10.1103/PhysRevLett.86.5377} {\bibfield  {journal} {\bibinfo  {journal}
  {Phys. Rev. Lett.}\ }\textbf {\bibinfo {volume} {86}},\ \bibinfo {pages}
  {5377} (\bibinfo {year} {2001})}\BibitemShut {NoStop}%
\bibitem [{\citenamefont {Christensen}\ \emph {et~al.}(2007)\citenamefont
  {Christensen}, \citenamefont {Ronnow}, \citenamefont {McMorrow},
  \citenamefont {Harrison}, \citenamefont {Perring}, \citenamefont {Enderle},
  \citenamefont {Coldea}, \citenamefont {Regnault},\ and\ \citenamefont
  {Aeppli}}]{2007Quantumdynamics}%
  \BibitemOpen
  \bibfield  {author} {\bibinfo {author} {\bibfnamefont {N.~B.}\ \bibnamefont
  {Christensen}}, \bibinfo {author} {\bibfnamefont {H.~M.}\ \bibnamefont
  {Ronnow}}, \bibinfo {author} {\bibfnamefont {D.~F.}\ \bibnamefont
  {McMorrow}}, \bibinfo {author} {\bibfnamefont {A.}~\bibnamefont {Harrison}},
  \bibinfo {author} {\bibfnamefont {T.~G.}\ \bibnamefont {Perring}}, \bibinfo
  {author} {\bibfnamefont {M.}~\bibnamefont {Enderle}}, \bibinfo {author}
  {\bibfnamefont {R.}~\bibnamefont {Coldea}}, \bibinfo {author} {\bibfnamefont
  {L.~P.}\ \bibnamefont {Regnault}}, \ and\ \bibinfo {author} {\bibfnamefont
  {G.}~\bibnamefont {Aeppli}},\ }\href
  {https://www.pnas.org/content/104/39/15264} {\bibfield  {journal} {\bibinfo
  {journal} {Proc. Natl. Acad. Sci. U. S. A.}\ }\textbf {\bibinfo {volume}
  {104}},\ \bibinfo {pages} {15264} (\bibinfo {year} {2007})}\BibitemShut
  {NoStop}%
\bibitem [{\citenamefont {Sala}\ \emph {et~al.}(2021)\citenamefont {Sala},
  \citenamefont {Stone}, \citenamefont {Rai}, \citenamefont {May},
  \citenamefont {Laurell}, \citenamefont {Garlea}, \citenamefont {Butch},
  \citenamefont {Lumsden}, \citenamefont {Ehlers}, \citenamefont {Pokharel},
  \citenamefont {Podlesnyak}, \citenamefont {Mandrus}, \citenamefont {Parker},
  \citenamefont {Okamoto}, \citenamefont {Hal\'{a}sz},\ and\ \citenamefont
  {Christianson}}]{2021Van}%
  \BibitemOpen
  \bibfield  {author} {\bibinfo {author} {\bibfnamefont {G.}~\bibnamefont
  {Sala}}, \bibinfo {author} {\bibfnamefont {M.~B.}\ \bibnamefont {Stone}},
  \bibinfo {author} {\bibfnamefont {B.~K.}\ \bibnamefont {Rai}}, \bibinfo
  {author} {\bibfnamefont {A.~F.}\ \bibnamefont {May}}, \bibinfo {author}
  {\bibfnamefont {P.}~\bibnamefont {Laurell}}, \bibinfo {author} {\bibfnamefont
  {V.~O.}\ \bibnamefont {Garlea}}, \bibinfo {author} {\bibfnamefont {N.~P.}\
  \bibnamefont {Butch}}, \bibinfo {author} {\bibfnamefont {M.~D.}\ \bibnamefont
  {Lumsden}}, \bibinfo {author} {\bibfnamefont {G.}~\bibnamefont {Ehlers}},
  \bibinfo {author} {\bibfnamefont {G.}~\bibnamefont {Pokharel}}, \bibinfo
  {author} {\bibfnamefont {A.}~\bibnamefont {Podlesnyak}}, \bibinfo {author}
  {\bibfnamefont {D.}~\bibnamefont {Mandrus}}, \bibinfo {author} {\bibfnamefont
  {D.~S.}\ \bibnamefont {Parker}}, \bibinfo {author} {\bibfnamefont
  {S.}~\bibnamefont {Okamoto}}, \bibinfo {author} {\bibfnamefont {G.~B.}\
  \bibnamefont {Hal\'{a}sz}}, \ and\ \bibinfo {author} {\bibfnamefont {A.~D.}\
  \bibnamefont {Christianson}},\ }\href {\doibase
  https://doi.org/10.1038/s41467-020-20335-5} {\bibfield  {journal} {\bibinfo
  {journal} {Nat. Commun.}\ }\textbf {\bibinfo {volume} {12}},\ \bibinfo
  {pages} {171} (\bibinfo {year} {2021})}\BibitemShut {NoStop}%
\bibitem [{\citenamefont {Wessler}\ \emph {et~al.}(2020)\citenamefont
  {Wessler}, \citenamefont {Roessli}, \citenamefont {Kr\"{a}mer}, \citenamefont
  {Delley}, \citenamefont {Waldmann}, \citenamefont {Keller}, \citenamefont
  {Cheptiakov}, \citenamefont {Braun},\ and\ \citenamefont
  {Kenzelmann}}]{npjqm.5.85}%
  \BibitemOpen
  \bibfield  {author} {\bibinfo {author} {\bibfnamefont {C.}~\bibnamefont
  {Wessler}}, \bibinfo {author} {\bibfnamefont {B.}~\bibnamefont {Roessli}},
  \bibinfo {author} {\bibfnamefont {K.~W.}\ \bibnamefont {Kr\"{a}mer}},
  \bibinfo {author} {\bibfnamefont {B.}~\bibnamefont {Delley}}, \bibinfo
  {author} {\bibfnamefont {O.}~\bibnamefont {Waldmann}}, \bibinfo {author}
  {\bibfnamefont {L.}~\bibnamefont {Keller}}, \bibinfo {author} {\bibfnamefont
  {D.}~\bibnamefont {Cheptiakov}}, \bibinfo {author} {\bibfnamefont {H.~B.}\
  \bibnamefont {Braun}}, \ and\ \bibinfo {author} {\bibfnamefont
  {M.}~\bibnamefont {Kenzelmann}},\ }\href {\doibase
  10.1038/s41535-020-00287-1} {\bibfield  {journal} {\bibinfo  {journal} {npj
  Quantum Mater.}\ }\textbf {\bibinfo {volume} {5}},\ \bibinfo {pages} {85}
  (\bibinfo {year} {2020})}\BibitemShut {NoStop}%
\bibitem [{\citenamefont {Mulder}\ \emph {et~al.}(2010)\citenamefont {Mulder},
  \citenamefont {Ganesh}, \citenamefont {Capriotti},\ and\ \citenamefont
  {Paramekanti}}]{PhysRevB.81.214419}%
  \BibitemOpen
  \bibfield  {author} {\bibinfo {author} {\bibfnamefont {A.}~\bibnamefont
  {Mulder}}, \bibinfo {author} {\bibfnamefont {R.}~\bibnamefont {Ganesh}},
  \bibinfo {author} {\bibfnamefont {L.}~\bibnamefont {Capriotti}}, \ and\
  \bibinfo {author} {\bibfnamefont {A.}~\bibnamefont {Paramekanti}},\ }\href
  {\doibase 10.1103/PhysRevB.81.214419} {\bibfield  {journal} {\bibinfo
  {journal} {Phys. Rev. B}\ }\textbf {\bibinfo {volume} {81}},\ \bibinfo
  {pages} {214419} (\bibinfo {year} {2010})}\BibitemShut {NoStop}%
\bibitem [{\citenamefont {Oitmaa}\ and\ \citenamefont
  {Singh}(2011)}]{PhysRevB.84.094424}%
  \BibitemOpen
  \bibfield  {author} {\bibinfo {author} {\bibfnamefont {J.}~\bibnamefont
  {Oitmaa}}\ and\ \bibinfo {author} {\bibfnamefont {R.~R.~P.}\ \bibnamefont
  {Singh}},\ }\href {\doibase 10.1103/PhysRevB.84.094424} {\bibfield  {journal}
  {\bibinfo  {journal} {Phys. Rev. B}\ }\textbf {\bibinfo {volume} {84}},\
  \bibinfo {pages} {094424} (\bibinfo {year} {2011})}\BibitemShut {NoStop}%
\bibitem [{\citenamefont {Farnell}\ \emph {et~al.}(2011)\citenamefont
  {Farnell}, \citenamefont {Bishop}, \citenamefont {Li}, \citenamefont
  {Richter},\ and\ \citenamefont {Campbell}}]{PhysRevB.84.012403}%
  \BibitemOpen
  \bibfield  {author} {\bibinfo {author} {\bibfnamefont {D.~J.~J.}\
  \bibnamefont {Farnell}}, \bibinfo {author} {\bibfnamefont {R.~F.}\
  \bibnamefont {Bishop}}, \bibinfo {author} {\bibfnamefont {P.~H.~Y.}\
  \bibnamefont {Li}}, \bibinfo {author} {\bibfnamefont {J.}~\bibnamefont
  {Richter}}, \ and\ \bibinfo {author} {\bibfnamefont {C.~E.}\ \bibnamefont
  {Campbell}},\ }\href {\doibase 10.1103/PhysRevB.84.012403} {\bibfield
  {journal} {\bibinfo  {journal} {Phys. Rev. B}\ }\textbf {\bibinfo {volume}
  {84}},\ \bibinfo {pages} {012403} (\bibinfo {year} {2011})}\BibitemShut
  {NoStop}%
\bibitem [{\citenamefont {Li}\ \emph {et~al.}(2014)\citenamefont {Li},
  \citenamefont {Bishop},\ and\ \citenamefont {Campbell}}]{PhysRevB.89.220408}%
  \BibitemOpen
  \bibfield  {author} {\bibinfo {author} {\bibfnamefont {P.~H.~Y.}\
  \bibnamefont {Li}}, \bibinfo {author} {\bibfnamefont {R.~F.}\ \bibnamefont
  {Bishop}}, \ and\ \bibinfo {author} {\bibfnamefont {C.~E.}\ \bibnamefont
  {Campbell}},\ }\href {\doibase 10.1103/PhysRevB.89.220408} {\bibfield
  {journal} {\bibinfo  {journal} {Phys. Rev. B}\ }\textbf {\bibinfo {volume}
  {89}},\ \bibinfo {pages} {220408} (\bibinfo {year} {2014})}\BibitemShut
  {NoStop}%
\bibitem [{\citenamefont {Ghorbani}\ \emph {et~al.}(2016)\citenamefont
  {Ghorbani}, \citenamefont {Shahbazi},\ and\ \citenamefont
  {Mosadeq}}]{2016Quantum}%
  \BibitemOpen
  \bibfield  {author} {\bibinfo {author} {\bibfnamefont {E.}~\bibnamefont
  {Ghorbani}}, \bibinfo {author} {\bibfnamefont {F.}~\bibnamefont {Shahbazi}},
  \ and\ \bibinfo {author} {\bibfnamefont {H.}~\bibnamefont {Mosadeq}},\ }\href
  {\doibase 10.1088/0953-8984/28/40/406001} {\bibfield  {journal} {\bibinfo
  {journal} {J. Phys.: Condens. Matter}\ }\textbf {\bibinfo {volume} {28}},\
  \bibinfo {pages} {406001} (\bibinfo {year} {2016})}\BibitemShut {NoStop}%
\bibitem [{\citenamefont {Clark}\ \emph {et~al.}(2011)\citenamefont {Clark},
  \citenamefont {Abanin},\ and\ \citenamefont
  {Sondhi}}]{PhysRevLett.107.087204}%
  \BibitemOpen
  \bibfield  {author} {\bibinfo {author} {\bibfnamefont {B.~K.}\ \bibnamefont
  {Clark}}, \bibinfo {author} {\bibfnamefont {D.~A.}\ \bibnamefont {Abanin}}, \
  and\ \bibinfo {author} {\bibfnamefont {S.~L.}\ \bibnamefont {Sondhi}},\
  }\href {\doibase 10.1103/PhysRevLett.107.087204} {\bibfield  {journal}
  {\bibinfo  {journal} {Phys. Rev. Lett.}\ }\textbf {\bibinfo {volume} {107}},\
  \bibinfo {pages} {087204} (\bibinfo {year} {2011})}\BibitemShut {NoStop}%
\bibitem [{\citenamefont {Di~Ciolo}\ \emph {et~al.}(2014)\citenamefont
  {Di~Ciolo}, \citenamefont {Carrasquilla}, \citenamefont {Becca},
  \citenamefont {Rigol},\ and\ \citenamefont {Galitski}}]{PhysRevB.89.094413}%
  \BibitemOpen
  \bibfield  {author} {\bibinfo {author} {\bibfnamefont {A.}~\bibnamefont
  {Di~Ciolo}}, \bibinfo {author} {\bibfnamefont {J.}~\bibnamefont
  {Carrasquilla}}, \bibinfo {author} {\bibfnamefont {F.}~\bibnamefont {Becca}},
  \bibinfo {author} {\bibfnamefont {M.}~\bibnamefont {Rigol}}, \ and\ \bibinfo
  {author} {\bibfnamefont {V.}~\bibnamefont {Galitski}},\ }\href {\doibase
  10.1103/PhysRevB.89.094413} {\bibfield  {journal} {\bibinfo  {journal} {Phys.
  Rev. B}\ }\textbf {\bibinfo {volume} {89}},\ \bibinfo {pages} {094413}
  (\bibinfo {year} {2014})}\BibitemShut {NoStop}%
\bibitem [{\citenamefont {Ferrari}\ \emph {et~al.}(2017)\citenamefont
  {Ferrari}, \citenamefont {Bieri},\ and\ \citenamefont
  {Becca}}]{PhysRevB.96.104401}%
  \BibitemOpen
  \bibfield  {author} {\bibinfo {author} {\bibfnamefont {F.}~\bibnamefont
  {Ferrari}}, \bibinfo {author} {\bibfnamefont {S.}~\bibnamefont {Bieri}}, \
  and\ \bibinfo {author} {\bibfnamefont {F.}~\bibnamefont {Becca}},\ }\href
  {\doibase 10.1103/PhysRevB.96.104401} {\bibfield  {journal} {\bibinfo
  {journal} {Phys. Rev. B}\ }\textbf {\bibinfo {volume} {96}},\ \bibinfo
  {pages} {104401} (\bibinfo {year} {2017})}\BibitemShut {NoStop}%
\bibitem [{\citenamefont {Fouet}\ \emph {et~al.}(2001)\citenamefont {Fouet},
  \citenamefont {Sindzingre},\ and\ \citenamefont {Lhuillier}}]{2001An}%
  \BibitemOpen
  \bibfield  {author} {\bibinfo {author} {\bibfnamefont {J.~B.}\ \bibnamefont
  {Fouet}}, \bibinfo {author} {\bibfnamefont {P.}~\bibnamefont {Sindzingre}}, \
  and\ \bibinfo {author} {\bibfnamefont {C.}~\bibnamefont {Lhuillier}},\ }\href
  {\doibase 10.1007/s100510170273} {\bibfield  {journal} {\bibinfo  {journal}
  {Eur. Phys. J. B}\ }\textbf {\bibinfo {volume} {20}},\ \bibinfo {pages} {241}
  (\bibinfo {year} {2001})}\BibitemShut {NoStop}%
\bibitem [{\citenamefont {Albuquerque}\ \emph {et~al.}(2011)\citenamefont
  {Albuquerque}, \citenamefont {Schwandt}, \citenamefont {Het\'enyi},
  \citenamefont {Capponi}, \citenamefont {Mambrini},\ and\ \citenamefont
  {L\"auchli}}]{PhysRevB.84.024406}%
  \BibitemOpen
  \bibfield  {author} {\bibinfo {author} {\bibfnamefont {A.~F.}\ \bibnamefont
  {Albuquerque}}, \bibinfo {author} {\bibfnamefont {D.}~\bibnamefont
  {Schwandt}}, \bibinfo {author} {\bibfnamefont {B.}~\bibnamefont {Het\'enyi}},
  \bibinfo {author} {\bibfnamefont {S.}~\bibnamefont {Capponi}}, \bibinfo
  {author} {\bibfnamefont {M.}~\bibnamefont {Mambrini}}, \ and\ \bibinfo
  {author} {\bibfnamefont {A.~M.}\ \bibnamefont {L\"auchli}},\ }\href {\doibase
  10.1103/PhysRevB.84.024406} {\bibfield  {journal} {\bibinfo  {journal} {Phys.
  Rev. B}\ }\textbf {\bibinfo {volume} {84}},\ \bibinfo {pages} {024406}
  (\bibinfo {year} {2011})}\BibitemShut {NoStop}%
\bibitem [{\citenamefont {Ganesh}\ \emph {et~al.}(2013)\citenamefont {Ganesh},
  \citenamefont {van~den Brink},\ and\ \citenamefont
  {Nishimoto}}]{PhysRevLett.110.127203}%
  \BibitemOpen
  \bibfield  {author} {\bibinfo {author} {\bibfnamefont {R.}~\bibnamefont
  {Ganesh}}, \bibinfo {author} {\bibfnamefont {J.}~\bibnamefont {van~den
  Brink}}, \ and\ \bibinfo {author} {\bibfnamefont {S.}~\bibnamefont
  {Nishimoto}},\ }\href {\doibase 10.1103/PhysRevLett.110.127203} {\bibfield
  {journal} {\bibinfo  {journal} {Phys. Rev. Lett.}\ }\textbf {\bibinfo
  {volume} {110}},\ \bibinfo {pages} {127203} (\bibinfo {year}
  {2013})}\BibitemShut {NoStop}%
\bibitem [{\citenamefont {Zhu}\ \emph {et~al.}(2013)\citenamefont {Zhu},
  \citenamefont {Huse},\ and\ \citenamefont {White}}]{PhysRevLett.110.127205}%
  \BibitemOpen
  \bibfield  {author} {\bibinfo {author} {\bibfnamefont {Z.}~\bibnamefont
  {Zhu}}, \bibinfo {author} {\bibfnamefont {D.~A.}\ \bibnamefont {Huse}}, \
  and\ \bibinfo {author} {\bibfnamefont {S.~R.}\ \bibnamefont {White}},\ }\href
  {\doibase 10.1103/PhysRevLett.110.127205} {\bibfield  {journal} {\bibinfo
  {journal} {Phys. Rev. Lett.}\ }\textbf {\bibinfo {volume} {110}},\ \bibinfo
  {pages} {127205} (\bibinfo {year} {2013})}\BibitemShut {NoStop}%
\bibitem [{\citenamefont {Gong}\ \emph {et~al.}(2013)\citenamefont {Gong},
  \citenamefont {Sheng}, \citenamefont {Motrunich},\ and\ \citenamefont
  {Fisher}}]{PhysRevB.88.165138}%
  \BibitemOpen
  \bibfield  {author} {\bibinfo {author} {\bibfnamefont {S.-S.}\ \bibnamefont
  {Gong}}, \bibinfo {author} {\bibfnamefont {D.~N.}\ \bibnamefont {Sheng}},
  \bibinfo {author} {\bibfnamefont {O.~I.}\ \bibnamefont {Motrunich}}, \ and\
  \bibinfo {author} {\bibfnamefont {M.~P.~A.}\ \bibnamefont {Fisher}},\ }\href
  {\doibase 10.1103/PhysRevB.88.165138} {\bibfield  {journal} {\bibinfo
  {journal} {Phys. Rev. B}\ }\textbf {\bibinfo {volume} {88}},\ \bibinfo
  {pages} {165138} (\bibinfo {year} {2013})}\BibitemShut {NoStop}%
\bibitem [{\citenamefont {Ferrari}\ and\ \citenamefont
  {Becca}(2020)}]{ferrari2020dynamical}%
  \BibitemOpen
  \bibfield  {author} {\bibinfo {author} {\bibfnamefont {F.}~\bibnamefont
  {Ferrari}}\ and\ \bibinfo {author} {\bibfnamefont {F.}~\bibnamefont
  {Becca}},\ }\href {\doibase 10.1088/1361-648X/ab7f6e} {\bibfield  {journal}
  {\bibinfo  {journal} {J. Phys.: Condens. Matter}\ }\textbf {\bibinfo {volume}
  {32}},\ \bibinfo {pages} {274003} (\bibinfo {year} {2020})}\BibitemShut
  {NoStop}%
\bibitem [{\citenamefont {Dong}\ \emph {et~al.}(2021)\citenamefont {Dong},
  \citenamefont {Wang}, \citenamefont {Gu},\ and\ \citenamefont
  {Li}}]{PhysRevB.104.L180406}%
  \BibitemOpen
  \bibfield  {author} {\bibinfo {author} {\bibfnamefont {Z.-Y.}\ \bibnamefont
  {Dong}}, \bibinfo {author} {\bibfnamefont {W.}~\bibnamefont {Wang}}, \bibinfo
  {author} {\bibfnamefont {Z.-L.}\ \bibnamefont {Gu}}, \ and\ \bibinfo {author}
  {\bibfnamefont {J.-X.}\ \bibnamefont {Li}},\ }\href {\doibase
  10.1103/PhysRevB.104.L180406} {\bibfield  {journal} {\bibinfo  {journal}
  {Phys. Rev. B}\ }\textbf {\bibinfo {volume} {104}},\ \bibinfo {pages}
  {L180406} (\bibinfo {year} {2021})}\BibitemShut {NoStop}%
\bibitem [{\citenamefont {Merino}\ and\ \citenamefont
  {Ralko}(2018)}]{PhysRevB.97.205112}%
  \BibitemOpen
  \bibfield  {author} {\bibinfo {author} {\bibfnamefont {J.}~\bibnamefont
  {Merino}}\ and\ \bibinfo {author} {\bibfnamefont {A.}~\bibnamefont {Ralko}},\
  }\href {\doibase 10.1103/PhysRevB.97.205112} {\bibfield  {journal} {\bibinfo
  {journal} {Phys. Rev. B}\ }\textbf {\bibinfo {volume} {97}},\ \bibinfo
  {pages} {205112} (\bibinfo {year} {2018})}\BibitemShut {NoStop}%
\bibitem [{\citenamefont {S\'{e}n\'{e}chal}\ \emph {et~al.}(2000)\citenamefont
  {S\'{e}n\'{e}chal}, \citenamefont {Perez},\ and\ \citenamefont
  {Pioro-Ladri\`{e}re}}]{Phys.Rev.Lett.84.522}%
  \BibitemOpen
  \bibfield  {author} {\bibinfo {author} {\bibfnamefont {D.}~\bibnamefont
  {S\'{e}n\'{e}chal}}, \bibinfo {author} {\bibfnamefont {D.}~\bibnamefont
  {Perez}}, \ and\ \bibinfo {author} {\bibfnamefont {M.}~\bibnamefont
  {Pioro-Ladri\`{e}re}},\ }\href {\doibase 10.1103/PhysRevLett.84.522}
  {\bibfield  {journal} {\bibinfo  {journal} {Phys. Rev. Lett.}\ }\textbf
  {\bibinfo {volume} {84}},\ \bibinfo {pages} {522} (\bibinfo {year}
  {2000})}\BibitemShut {NoStop}%
\bibitem [{\citenamefont {Zacher}\ \emph {et~al.}(2000)\citenamefont {Zacher},
  \citenamefont {Eder}, \citenamefont {Arrigoni},\ and\ \citenamefont
  {Hanke}}]{PhysRevLett.85.2585}%
  \BibitemOpen
  \bibfield  {author} {\bibinfo {author} {\bibfnamefont {M.~G.}\ \bibnamefont
  {Zacher}}, \bibinfo {author} {\bibfnamefont {R.}~\bibnamefont {Eder}},
  \bibinfo {author} {\bibfnamefont {E.}~\bibnamefont {Arrigoni}}, \ and\
  \bibinfo {author} {\bibfnamefont {W.}~\bibnamefont {Hanke}},\ }\href
  {\doibase 10.1103/PhysRevLett.85.2585} {\bibfield  {journal} {\bibinfo
  {journal} {Phys. Rev. Lett.}\ }\textbf {\bibinfo {volume} {85}},\ \bibinfo
  {pages} {2585} (\bibinfo {year} {2000})}\BibitemShut {NoStop}%
\bibitem [{\citenamefont {S\'{e}n\'{e}chal}\ and\ \citenamefont
  {Tremblay}(2004)}]{PhysRevLett.92.126401}%
  \BibitemOpen
  \bibfield  {author} {\bibinfo {author} {\bibfnamefont {D.}~\bibnamefont
  {S\'{e}n\'{e}chal}}\ and\ \bibinfo {author} {\bibfnamefont {A.-M.~S.}\
  \bibnamefont {Tremblay}},\ }\href {\doibase 10.1103/PhysRevLett.92.126401}
  {\bibfield  {journal} {\bibinfo  {journal} {Phys. Rev. Lett.}\ }\textbf
  {\bibinfo {volume} {92}},\ \bibinfo {pages} {126401} (\bibinfo {year}
  {2004})}\BibitemShut {NoStop}%
\bibitem [{\citenamefont {Yu}\ \emph {et~al.}(2011)\citenamefont {Yu},
  \citenamefont {Xie},\ and\ \citenamefont {Li}}]{PhysRevLett.107.010401}%
  \BibitemOpen
  \bibfield  {author} {\bibinfo {author} {\bibfnamefont {S.-L.}\ \bibnamefont
  {Yu}}, \bibinfo {author} {\bibfnamefont {X.~C.}\ \bibnamefont {Xie}}, \ and\
  \bibinfo {author} {\bibfnamefont {J.-X.}\ \bibnamefont {Li}},\ }\href
  {\doibase 10.1103/PhysRevLett.107.010401} {\bibfield  {journal} {\bibinfo
  {journal} {Phys. Rev. Lett.}\ }\textbf {\bibinfo {volume} {107}},\ \bibinfo
  {pages} {010401} (\bibinfo {year} {2011})}\BibitemShut {NoStop}%
\bibitem [{\citenamefont {Kang}\ \emph {et~al.}(2011)\citenamefont {Kang},
  \citenamefont {Yu}, \citenamefont {Xiang},\ and\ \citenamefont
  {Li}}]{PhysRevB.84.064520}%
  \BibitemOpen
  \bibfield  {author} {\bibinfo {author} {\bibfnamefont {J.}~\bibnamefont
  {Kang}}, \bibinfo {author} {\bibfnamefont {S.-L.}\ \bibnamefont {Yu}},
  \bibinfo {author} {\bibfnamefont {T.}~\bibnamefont {Xiang}}, \ and\ \bibinfo
  {author} {\bibfnamefont {J.-X.}\ \bibnamefont {Li}},\ }\href {\doibase
  10.1103/PhysRevB.84.064520} {\bibfield  {journal} {\bibinfo  {journal} {Phys.
  Rev. B}\ }\textbf {\bibinfo {volume} {84}},\ \bibinfo {pages} {064520}
  (\bibinfo {year} {2011})}\BibitemShut {NoStop}%
\bibitem [{\citenamefont {Yu}\ and\ \citenamefont
  {Li}(2012)}]{PhysRevB.85.144402}%
  \BibitemOpen
  \bibfield  {author} {\bibinfo {author} {\bibfnamefont {S.-L.}\ \bibnamefont
  {Yu}}\ and\ \bibinfo {author} {\bibfnamefont {J.-X.}\ \bibnamefont {Li}},\
  }\href {\doibase 10.1103/PhysRevB.85.144402} {\bibfield  {journal} {\bibinfo
  {journal} {Phys. Rev. B}\ }\textbf {\bibinfo {volume} {85}},\ \bibinfo
  {pages} {144402} (\bibinfo {year} {2012})}\BibitemShut {NoStop}%
\bibitem [{\citenamefont {Matsubara}\ and\ \citenamefont
  {Matsuda}(1956)}]{ProgTheorPhys.16.569}%
  \BibitemOpen
  \bibfield  {author} {\bibinfo {author} {\bibfnamefont {T.}~\bibnamefont
  {Matsubara}}\ and\ \bibinfo {author} {\bibfnamefont {H.}~\bibnamefont
  {Matsuda}},\ }\href {\doibase 10.1143/PTP.16.569} {\bibfield  {journal}
  {\bibinfo  {journal} {Prog. Theor. Phys.}\ }\textbf {\bibinfo {volume}
  {16}},\ \bibinfo {pages} {569} (\bibinfo {year} {1956})}\BibitemShut
  {NoStop}%
\bibitem [{\citenamefont {Batyev}\ and\ \citenamefont
  {Braginskii}(1984)}]{Sov.Phys.JETP.60.781}%
  \BibitemOpen
  \bibfield  {author} {\bibinfo {author} {\bibfnamefont {E.~G.}\ \bibnamefont
  {Batyev}}\ and\ \bibinfo {author} {\bibfnamefont {L.~S.}\ \bibnamefont
  {Braginskii}},\ }\href
  {http://www.jetp.ac.ru/cgi-bin/e/index/r/87/4/p1361?a=list} {\bibfield
  {journal} {\bibinfo  {journal} {Sov. Phys. JETP}\ }\textbf {\bibinfo {volume}
  {60}},\ \bibinfo {pages} {781} (\bibinfo {year} {1984})}\BibitemShut
  {NoStop}%
\bibitem [{\citenamefont {Dagotto}(1994)}]{RevModPhys.66.763}%
  \BibitemOpen
  \bibfield  {author} {\bibinfo {author} {\bibfnamefont {E.}~\bibnamefont
  {Dagotto}},\ }\href {\doibase 10.1103/RevModPhys.66.763} {\bibfield
  {journal} {\bibinfo  {journal} {Rev. Mod. Phys.}\ }\textbf {\bibinfo {volume}
  {66}},\ \bibinfo {pages} {763} (\bibinfo {year} {1994})}\BibitemShut
  {NoStop}%
\bibitem [{\citenamefont {Jakli\ifmmode~\check{c}\else \v{c}\fi{}}\ and\
  \citenamefont {Prelov\ifmmode~\check{s}\else
  \v{s}\fi{}ek}(1994)}]{PhysRevB.49.5065}%
  \BibitemOpen
  \bibfield  {author} {\bibinfo {author} {\bibfnamefont {J.}~\bibnamefont
  {Jakli\ifmmode~\check{c}\else \v{c}\fi{}}}\ and\ \bibinfo {author}
  {\bibfnamefont {P.}~\bibnamefont {Prelov\ifmmode~\check{s}\else
  \v{s}\fi{}ek}},\ }\href {\doibase 10.1103/PhysRevB.49.5065} {\bibfield
  {journal} {\bibinfo  {journal} {Phys. Rev. B}\ }\textbf {\bibinfo {volume}
  {49}},\ \bibinfo {pages} {5065} (\bibinfo {year} {1994})}\BibitemShut
  {NoStop}%
\bibitem [{\citenamefont {Majumdar}(2010)}]{PhysRevB.82.144407}%
  \BibitemOpen
  \bibfield  {author} {\bibinfo {author} {\bibfnamefont {K.}~\bibnamefont
  {Majumdar}},\ }\href {\doibase 10.1103/PhysRevB.82.144407} {\bibfield
  {journal} {\bibinfo  {journal} {Phys. Rev. B}\ }\textbf {\bibinfo {volume}
  {82}},\ \bibinfo {pages} {144407} (\bibinfo {year} {2010})}\BibitemShut
  {NoStop}%
\bibitem [{\citenamefont {Ma}\ \emph {et~al.}(2018)\citenamefont {Ma},
  \citenamefont {Sun}, \citenamefont {You}, \citenamefont {Xu}, \citenamefont
  {Vishwanath}, \citenamefont {Sandvik},\ and\ \citenamefont
  {Meng}}]{PhysRevB.98.174421}%
  \BibitemOpen
  \bibfield  {author} {\bibinfo {author} {\bibfnamefont {N.}~\bibnamefont
  {Ma}}, \bibinfo {author} {\bibfnamefont {G.-Y.}\ \bibnamefont {Sun}},
  \bibinfo {author} {\bibfnamefont {Y.-Z.}\ \bibnamefont {You}}, \bibinfo
  {author} {\bibfnamefont {C.}~\bibnamefont {Xu}}, \bibinfo {author}
  {\bibfnamefont {A.}~\bibnamefont {Vishwanath}}, \bibinfo {author}
  {\bibfnamefont {A.~W.}\ \bibnamefont {Sandvik}}, \ and\ \bibinfo {author}
  {\bibfnamefont {Z.~Y.}\ \bibnamefont {Meng}},\ }\href {\doibase
  10.1103/PhysRevB.98.174421} {\bibfield  {journal} {\bibinfo  {journal} {Phys.
  Rev. B}\ }\textbf {\bibinfo {volume} {98}},\ \bibinfo {pages} {174421}
  (\bibinfo {year} {2018})}\BibitemShut {NoStop}%
\bibitem [{\citenamefont {Senthil}\ \emph
  {et~al.}(2004{\natexlab{a}})\citenamefont {Senthil}, \citenamefont {Balents},
  \citenamefont {Sachdev}, \citenamefont {Vishwanath},\ and\ \citenamefont
  {Fisher}}]{PhysRevB.70.144407}%
  \BibitemOpen
  \bibfield  {author} {\bibinfo {author} {\bibfnamefont {T.}~\bibnamefont
  {Senthil}}, \bibinfo {author} {\bibfnamefont {L.}~\bibnamefont {Balents}},
  \bibinfo {author} {\bibfnamefont {S.}~\bibnamefont {Sachdev}}, \bibinfo
  {author} {\bibfnamefont {A.}~\bibnamefont {Vishwanath}}, \ and\ \bibinfo
  {author} {\bibfnamefont {M.~P.~A.}\ \bibnamefont {Fisher}},\ }\href {\doibase
  10.1103/PhysRevB.70.144407} {\bibfield  {journal} {\bibinfo  {journal} {Phys.
  Rev. B}\ }\textbf {\bibinfo {volume} {70}},\ \bibinfo {pages} {144407}
  (\bibinfo {year} {2004}{\natexlab{a}})}\BibitemShut {NoStop}%
\bibitem [{\citenamefont {Senthil}\ \emph
  {et~al.}(2004{\natexlab{b}})\citenamefont {Senthil}, \citenamefont
  {Vishwanath}, \citenamefont {Balents}, \citenamefont {Sachdev},\ and\
  \citenamefont {Fisher}}]{Senthil2004}%
  \BibitemOpen
  \bibfield  {author} {\bibinfo {author} {\bibfnamefont {T.}~\bibnamefont
  {Senthil}}, \bibinfo {author} {\bibfnamefont {A.}~\bibnamefont {Vishwanath}},
  \bibinfo {author} {\bibfnamefont {L.}~\bibnamefont {Balents}}, \bibinfo
  {author} {\bibfnamefont {S.}~\bibnamefont {Sachdev}}, \ and\ \bibinfo
  {author} {\bibfnamefont {M.}~\bibnamefont {Fisher}},\ }\href
  {https://www.science.org/doi/10.1126/science.1091806} {\bibfield  {journal}
  {\bibinfo  {journal} {Science}\ }\textbf {\bibinfo {volume} {303}} (\bibinfo
  {year} {2004}{\natexlab{b}})}\BibitemShut {NoStop}%
\bibitem [{\citenamefont {Zhitomirsky}\ and\ \citenamefont
  {Chernyshev}(2013)}]{RevModPhys.85.219}%
  \BibitemOpen
  \bibfield  {author} {\bibinfo {author} {\bibfnamefont {M.~E.}\ \bibnamefont
  {Zhitomirsky}}\ and\ \bibinfo {author} {\bibfnamefont {A.~L.}\ \bibnamefont
  {Chernyshev}},\ }\href {\doibase 10.1103/RevModPhys.85.219} {\bibfield
  {journal} {\bibinfo  {journal} {Rev. Mod. Phys.}\ }\textbf {\bibinfo {volume}
  {85}},\ \bibinfo {pages} {219} (\bibinfo {year} {2013})}\BibitemShut
  {NoStop}%
\bibitem [{\citenamefont {Winter}\ \emph {et~al.}(2017)\citenamefont {Winter},
  \citenamefont {Riedl}, \citenamefont {Maksimov}, \citenamefont {Chernyshev},
  \citenamefont {Honecker},\ and\ \citenamefont {Valent}}]{Nat.Commun.8.1152}%
  \BibitemOpen
  \bibfield  {author} {\bibinfo {author} {\bibfnamefont {S.~M.}\ \bibnamefont
  {Winter}}, \bibinfo {author} {\bibfnamefont {K.}~\bibnamefont {Riedl}},
  \bibinfo {author} {\bibfnamefont {P.~A.}\ \bibnamefont {Maksimov}}, \bibinfo
  {author} {\bibfnamefont {A.~L.}\ \bibnamefont {Chernyshev}}, \bibinfo
  {author} {\bibfnamefont {A.}~\bibnamefont {Honecker}}, \ and\ \bibinfo
  {author} {\bibfnamefont {R.}~\bibnamefont {Valent}},\ }\href {\doibase
  10.1038/s41467-017-01177-0} {\bibfield  {journal} {\bibinfo  {journal} {Nat.
  Commun.}\ }\textbf {\bibinfo {volume} {8}},\ \bibinfo {pages} {1152}
  (\bibinfo {year} {2017})}\BibitemShut {NoStop}%
\end{thebibliography}%

\end{document}